\RequirePackage{fix-cm}
\documentclass[twocolumn,final,epjc3]{svjour3}  
\usepackage{cite}
\RequirePackage{graphicx}
\usepackage{isotope} 
\usepackage{upgreek} 
\usepackage{hyperref}
\usepackage{cancel}
\usepackage[separate-uncertainty=true]{siunitx} 
\usepackage{color}
\usepackage{amsmath}
\usepackage{subfig}
\renewcommand \thesection {\Roman{section}}

\newcommand{\munster}{Institut f\"ur Kernphysik, Westf\"alische Wilhelms-Universit\"at M\"unster, M\"unster, Germany}

\journalname{Eur. Phys. J. C}
\begin{document}
\title{Design, construction and commissioning of a high-flow radon removal system for XENONnT}
\author{M.~Murra\thanksref{munster, murra,murra2}
\and
		D.~Schulte\thanksref{munster, schulte}
\and
		C.~Huhmann\thanksref{munster}
\and
		C.~Weinheimer\thanksref{munster}
		}	   
\thankstext{murra}{e-mail: michaelmurra@uni-muenster.de}
\thankstext{schulte}{e-mail: DennySchulte@uni-muenster.de}
\thankstext{murra2}{New address: Physics Department, Columbia University, New York, USA}

\institute{\munster \label{munster}}
%
%
%
%
%
\maketitle

\begin{abstract}
A high-flow radon removal system based on cryogenic distillation was developed and constructed to reduce radon-induced backgrounds in liquid xenon detectors for rare event searches such as XENONnT. A continuous purification of the XENONnT liquid xenon inventory of \SI{8.4}{tonnes} at process flows up to 71\,kg/h (200\,slpm) is required to achieve a radon reduction by a factor two for radon sources inside the detector. To reach such high flows, the distillation column's design features liquid xenon inlet and outlets along with novel custom-made bath-type heat exchangers with high liquefaction capabilities. The distillation process was designed using a modification of the McCabe-Thiele approach without a bottom product extraction. The thermodynamic concept is based on a Clausius-Rankine cooling cycle with phase-changing medium, in this case the xenon itself. To drastically reduce the external cooling power requirements, an energy efficient heat pump concept was developed applying a custom-made four cylinder magnetically-coupled piston pump as compressor. The distillation system was operated at thermodynamically stable conditions at a process flow of \linebreak \SI{91(2)}{kg/h} (($258\pm6$)\,slpm), 30\% over design. With this flow, a \isotope[222]{Rn} activity concentration $<1\,\upmu$Bq/kg is expected inside the XENONnT detector given the measured radon source distribution.

\end{abstract}

\section{Introduction}
\label{sec:Introduction}
Liquid xenon (LXe) is a well-established target material for rare event searches\,\cite{a,b,c,d,e,f,g,h}.
For XENONnT, located underground at the Laboratori Nazionali del Gran Sasso (LNGS), with 8.4 tonnes of xenon, \isotope[222]{Rn} is the main source of background for the dark matter WIMP search \cite{b}, while the radioactives \isotope[219]{Rn} and \isotope[220]{Rn} are less critical due to their short half-life.
Radon is continuously emanated from the detector materials into the target material xenon. Its half-life of \SI{3.8}{d} allows for a homogeneous mixing within the active detection volume. 
The beta decay of \isotope[214]{Pb} within the \isotope[222]{Rn} decay chain cannot be tagged by time coincidences and thus, creates leakage events into the dark matter region of interest.
Therefore, radon should be avoided in the first place as much as possible by dedicated material screening and selection \cite{Eman1,Eman2}. 
Emanation measurements reveal the location of different radon sources in the global detector system including the cryostat, but also the adjacent xenon handling systems like cryogenic and purification systems.

This makes it possible to create a radon reduction strategy with an online radon removal system (RRS). Such a system can be based on a charcoal tower \cite{xm, lz} or, in the case of XENONnT, based on a cryogenic distillation tower making use of the difference in vapor pressure between xenon and radon. Radon, as the less volatile component compared to xenon, will accumulate within the liquid xenon at the bottom of the distillation tower. Here, it is trapped until disintegration. Thus, no extraction of the enriched xenon is required making the operation xenon-loss-free.

\begin{sloppypar} Due to the large detector mass, a large xenon process flow is required for the RRS to reach reduction factors inside the detector of a factor two and above. Therefore, the XENONnT RRS features liquid xenon inlet and outlet flows requiring a special thermodynamic concept. This is based on a heat-pump cooling concept realized with radon-free heat exchangers and a compressor. The design, construction and the thermodynamic commissioning of this novel high-flow radon removal system for XENONnT is presented in this work. The radon removal performance is published in \cite{RAD} using XENONnT directly as a radon source and monitor.\end{sloppypar}

In section \ref{sec:radonremovalstrategy}, a radon removal strategy based on the XENONnT radon source distribution is developed. From this, the required design parameters for a radon distillation system can be derived. The design of the distillation tower is explained in section \ref{sec:McCabe} based on a modified McCabe-Thiele method without bottom product extraction. Section \ref{sec:radon_column} shows the thermodynamic concept based on a Clausius-Rankine cycle as well as the full experimental setup including different key components. The thermodynamic commissioning through an internal liquid xenon bypass valve is presented in section \ref{sec:commissioning} followed by a conclusion in section \ref{sec:conclusion}.
\section{Radon removal strategy}
\label{sec:radonremovalstrategy}
\begin{figure}[!h]
	\centering
    \includegraphics[width=\linewidth]{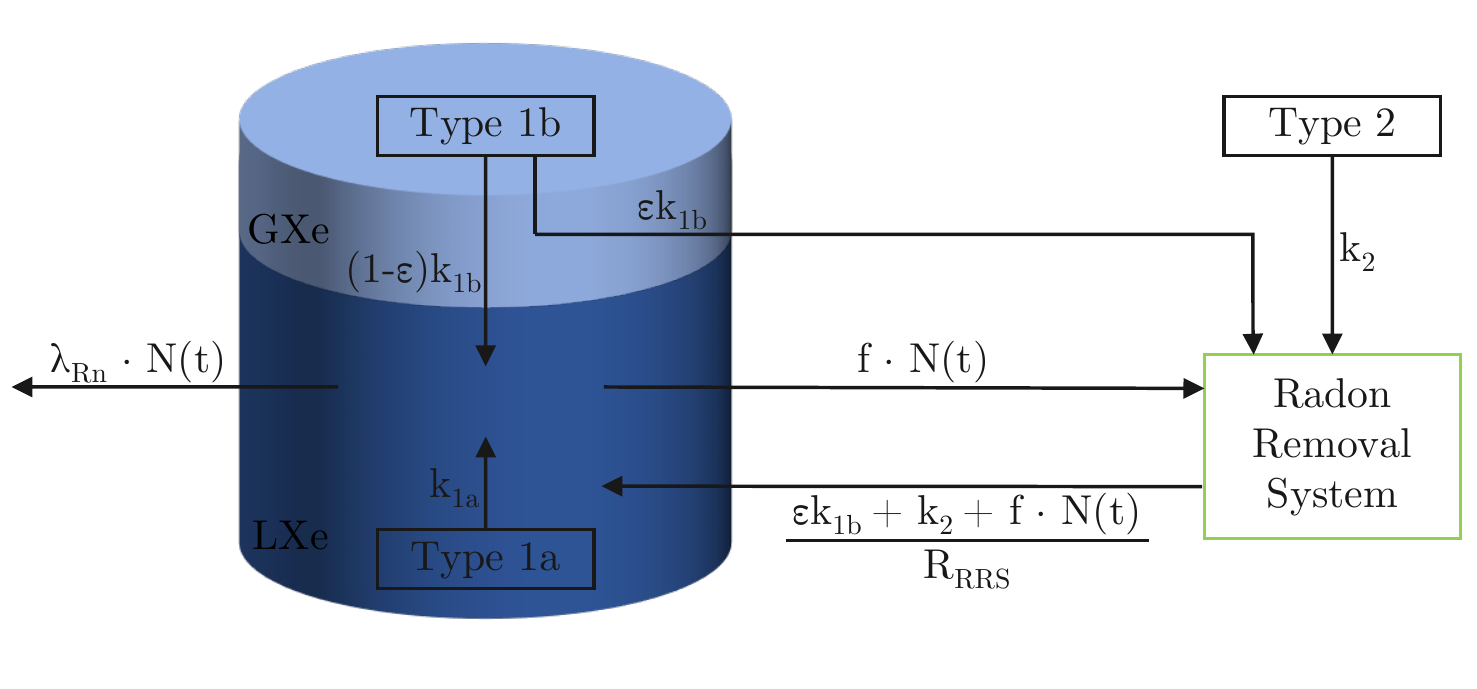}
 	\caption{Simplified scheme for a radon removal  strategy. Different radon sources (type 1a, type 1b, type 2) with regard to their location within the global system have to be considered and are indicated by their production rate $k_\mathrm{1a}$,  $k_\mathrm{1b}$ and $k_\mathrm{2}$.}
 	\label{fig:removalscheme}
\end{figure}

A radon removal strategy can be defined based on the radon source distribution within the global system. A simplified version is visualized in figure \ref{fig:removalscheme} including a detector with gaseous (GXe) and liquid phase as well as a radon removal system.

The radon emanation sources are divided in two types of sources, depending of their position in the global detector system with respect to the RRS: Sources that are labeled as type 1 are being emanated and flushed directly into the detector before reaching the removal system. They are subdivided into sources directly inside the LXe (type 1a) and sources within the GXe phase (type 1b) indicated as $k_\mathrm{1a}$ and $k_\mathrm{1b}$ in figure \ref{fig:removalscheme}, respectively. A radon source between the detector and the RRS is referred to as type 2. Consequently, type 2 radon is going into the RRS before reaching the LXe inside the detector ($k_\mathrm{2}$). 

The removal strategy is split in two ways: Firstly, radon is removed by continuously extracting a LXe flow from the detector, purify it with the RRS, and feed radon-depleted LXe back. Secondly, radon is removed by extracting xenon from the GXe phase before entering the detector's LXe phase. This way, type 1b sources can be converted into type 2 sources going directly to the RRS. The extraction efficiency for this process is defined as $\epsilon$.

The change in the number of radon particles inside the detector's LXe volume over time $\frac{dN(t)}{dt}$ can be expressed by the following differential equation taking all in-going and out-going radon atoms into account:
\begin{align}\label{eq:DGL}
\frac{dN(t)}{dt} = & \hspace{1mm}k_\mathrm{1a} + (1 - \epsilon) k_\mathrm{1b}\nonumber\\
&- \lambda_{\mathrm{Rn}} \cdot N(t) - f \cdot N(t)\\
&+ \frac{k_2 + \epsilon k_\mathrm{1b} + f \cdot N(t)}{R_\mathrm{RRS}}\nonumber.
\end{align}

Here, $\lambda_{\mathrm{Rn}} \cdot N(t)$ denotes the decay of \isotope[222]{Rn} with its decay constant $\lambda_{\mathrm{Rn}} = \SI{0.18}{1/d}$ and $f \cdot N(t)$ describes the effective radon particle flux leaving the detector. The effective circulation rate $f$ depends on the purification flow $F_\mathrm{Xe}$ as well as on the total amount of the xenon inventory $m_{\mathrm{Xe}}$ and is defined as
\begin{equation}
f = \frac{F_{\mathrm{Xe}}}{m_{\mathrm{Xe}}} = \frac{1}{\tau_{\textrm{ex}}}, 
\end{equation}

with $\tau_{\textrm{ex}}$ being the detector's LXe volume exchange time. The number of radon atoms that are not removed by the RRS are given by the last term of equation (\ref{eq:DGL}) and depends on the RRS's reduction factor $R_\mathrm{RRS}$. It is defined as the ratio of the radon concentration at the inlet feed $c_\mathrm{F}$ and at the purified outlet $c_\mathrm{D}$ with
\begin{equation}
R_\mathrm{RRS} \equiv \frac{c_{\mathrm{F}}}{c_{\mathrm{D}}}.
\end{equation}
Furthermore, $R_\mathrm{RRS}$ is assumed to be constant and independent from the radon concentration.

Equation (\ref{eq:DGL}) can be solved for the starting condition $N(t=0) = N_0$, where $N_0$ is the number of radon atoms before starting the removal:
\begin{equation}\label{eq:DGL_solution}
N(t) = \frac{K}{\Lambda} + \left(N_0 - \frac{K}{\Lambda}\right) \cdot e^{-\Lambda \cdot t},
\end{equation}
with
\begin{equation}
K = k_\mathrm{1a} + (1 - \epsilon) k_\mathrm{1b} + \frac{k_2 + \epsilon k_\mathrm{1b}}{R_\mathrm{RRS}}
,
\end{equation}
and
\begin{equation}
\Lambda = \left( \lambda_{\mathrm{Rn}} + f \cdot \left(1-\frac{1}{R_\mathrm{RRS}}\right)\right).
\end{equation}
For infinitely large times, equation (\ref{eq:DGL_solution}) simplifies to the equilibrium relation
\begin{align}
N_\mathrm{equi} &\overset{t\to \infty}{=} \frac{K}{\Lambda} = \frac{k_\mathrm{1a} + (1 - \epsilon) k_\mathrm{1b} + \frac{k_2 + \epsilon k_\mathrm{1b}}{R_\mathrm{RRS}}}{\lambda_{\mathrm{Rn}} + f \cdot (1-\frac{1}{R_\mathrm{RRS}})} \\
&\overset{R_\mathrm{RRS}\to \infty}{=}\frac{1}{\lambda_{\mathrm{Rn}} + f}\cdot {(k_\mathrm{1a} + (1 - \epsilon) k_\mathrm{1b})},
\end{align}
where the last step assumes an infinite reduction power $R_\mathrm{RRS}$. This shows that a perfect RRS can fully remove radon emanated directly by type 2 sources as well as type 1b sources converted into type 2. The effective flow $f$ is limiting the reduction of type 1 sources. The reduction inside the detector's LXe volume can be defined as
\begin{equation}
r(R_\mathrm{RRS},f, \epsilon) \equiv \frac{N_\mathrm{equi}(R_\mathrm{RRS}=1, f, \epsilon)}{N_\mathrm{equi}(R_\mathrm{RRS},f,\epsilon)}.
\end{equation}
For a highly efficient RRS ($R_\mathrm{RRS}\to \infty$) one finds
\begin{equation}
\label{eq:lxe_gxe_reduction}
r(R_\mathrm{RRS}\to \infty, f, \epsilon)
=\frac{\lambda_\mathrm{Rn} + f}{\lambda_{\mathrm{Rn}}}\cdot \frac{k_\mathrm{tot}}{k_\mathrm{1a} + (1 - \epsilon) k_\mathrm{1b}},
\end{equation}
with $k_\mathrm{tot}=k_\mathrm{1a}+k_\mathrm{1b}+k_\mathrm{2}$. Equation (\ref{eq:lxe_gxe_reduction}) is valid for the combined LXe and GXe modes.
The individual reduction capabilities $r_\mathrm{LXe}$ for the LXe only extraction ($f > 0$, $\epsilon = 0$) and $r_\mathrm{GXe}$ for the GXe only extraction ($f = 0$, $\epsilon > 0$) are given by
\begin{align}
r_\mathrm{LXe} &=r(R_\mathrm{RRS}\to \infty,f, \epsilon=0) \\ \notag
&=  \frac{\lambda_\mathrm{Rn} + f}{\lambda_{\mathrm{Rn}}}\cdot
 \frac{k_\mathrm{tot}}{k_\mathrm{1a} +  k_\mathrm{1b}},
\end{align}
and
\begin{align}
r_\mathrm{GXe} &=r(R_\mathrm{RRS}\to \infty,f=0, \epsilon) \\ \notag
&=  \frac{k_\mathrm{tot}}{k_\mathrm{1a} + (1 - \epsilon) k_\mathrm{1b}}.
\end{align}

For systems like XENONnT with $(k_\mathrm{1a} + k_\mathrm{1b}) \gg k_\mathrm{2}$, a reduction $r$ inside the detector by a factor two can be reached for $f =  \lambda_\mathrm{Rn}$, i.e. the RRS's purification flow needs to exchange a detector's LXe inventory with a time constant of $\tau_{\textrm{ex}} = \SI{5.5}{d}$. In the case of XENONnT, this translates to \SI{62}{kg/h} (\SI{175}{slpm}) for a xenon inventory of \SI{8.4}{t} and defines the required performance of the RRS for XENONnT. 

A further boost in reduction can be achieved by converting type 1b into type 2 sources, the GXe mode. An ideal extraction ($\epsilon = 1$) yields a maximum reduction factor $r$ of $2 \cdot \frac{k_\mathrm{tot}}{k_\mathrm{1a}}$ assuming $f =  \lambda_\mathrm{Rn}$.

In the following, the expected radon reduction within the XENONnT detector is evaluated: The dual-phase LXe Time Projection Chamber (LXe TPC) itself is housed in a cryostat referred to as the inner vessel. In order to keep the xenon of the detector system in a thermodynamic equilibrium, a cryogenic system (CRY) is required. Here, two pulse tube refrigerators and an emergency LN$_2$ cooling tower are connected to the inner vessel via a cryopipe to balance evaporated xenon due to external heat input by condensation. Two cable feedthrough vessels (Cables) connected to the cryopipe contain high voltage and signal cables of the PMTs inside the detector. Furthermore, two purification systems are used to clean the xenon: LXe from the detector is extracted and evaporated via a series of heat exchangers in order to be purified from electronegative impurities with the help of a gas purification system (GXe PUR). In parallel, a novel LXe purification loop (LXe PUR) allows to circulate the xenon inventory once a day through LXe filters removing electronegatives as well. While most of the xenon through the LXe PUR is returned to the cryopipe and subsequently to the detector, a fraction is guided to the RRS for radon reduction. All details about the XENONnT subsystems are summarized in Ref. \cite{Instr}.

The radon source locations in XENONnT are well known due to dedicated emanation measurements \cite{Eman2}. The values for the different sources are typically stated in units of a \isotope[222]{Rn} activity concentration per kilogram of xenon and are summarized in table \ref{tab:radonsources} assuming a \SI{8.4}{t} xenon inventory. Note that radon sources related to the RRS are only contributing to the total budget when the RRS is in operation. During this operation, the LXe fraction $\epsilon_{\mathrm{LP}}$ coming from the LXe PUR system into the RRS is given by $F_\mathrm{Xe}= \epsilon_\mathrm{LP}\cdot G_\mathrm{Xe}$
with the total LXe purification flow $G_\mathrm{Xe}$. Assuming a homogeneous radon distribution inside the LXe flow, $\epsilon_\mathrm{LP}$ corresponds to the fraction of the measured radon emanation from the LXe PUR in table\,\ref{tab:radonsources} that contributes as a type 2 source. In the following, $\epsilon_\mathrm{LP}$ was fixed to \num{0.2} by design.

The CRY system and the Cables have separated extraction ports, and thus, feature different extraction efficiencies $\epsilon_{\mathrm{CRY}}$ and $\epsilon_{\mathrm{C}}$, respectively, contributing to equation\,\ref{eq:DGL} with
\begin{align}
\epsilon k_{\mathrm{1b}} &= \epsilon_{\mathrm{CRY}} k_{\mathrm{CRY}}+\epsilon_{\mathrm{C}} k_{\mathrm{C}},\\ 
(1 -\epsilon) k_{\mathrm{1b}} &= (1 -\epsilon_{\mathrm{CRY}}) k_{\mathrm{CRY}}+(1 -\epsilon_{\mathrm{C}}) k_{\mathrm{C}}.
\end{align}
\begin{table}[htbp]
\caption{Radon source distribution in XENONnT assuming a \SI{8.4}{t} xenon inventory \cite{Eman2}.}
\centering
\begin{tabular*}{\columnwidth}{lccc}
\hline
 & System & Variable & Value [$\upmu$Bq/kg]\\ \hline
 Type 1a & LXe TPC & $k_{\mathrm{TPC}}$ & 1.11\\
         & Inner vessel & $k_{\mathrm{IV}}$ & 0.30 \\
         & GXe PUR & $k_{\mathrm{GP}}$ & 0.21\\
         & LXe PUR & $k_{\mathrm{LP,I}}$ & $(1-\epsilon_{\mathrm{LP}})\times$ 0.43\\ 
         & RRS-outlet & $k_{\mathrm{RRS,I}}$ & 0.19\\\hline 
 Type 1b & CRY & $k_{\mathrm{CRY}}$ & 1.36\\
         & Cables & $k_{\mathrm{C}}$ & 0.85\\\hline 
 Type 2 & RRS-inlet & $k_{\mathrm{RRS,II}}$ & 0.3\\
         & LXe PUR & $k_{\mathrm{LP,II}}$ & $\epsilon_{\mathrm{LP}}\times$ 0.43\\ \hline 
\end{tabular*}
\label{tab:radonsources}
\end{table}

\begin{figure}[!h]
	\centering
    \includegraphics[width=\linewidth]{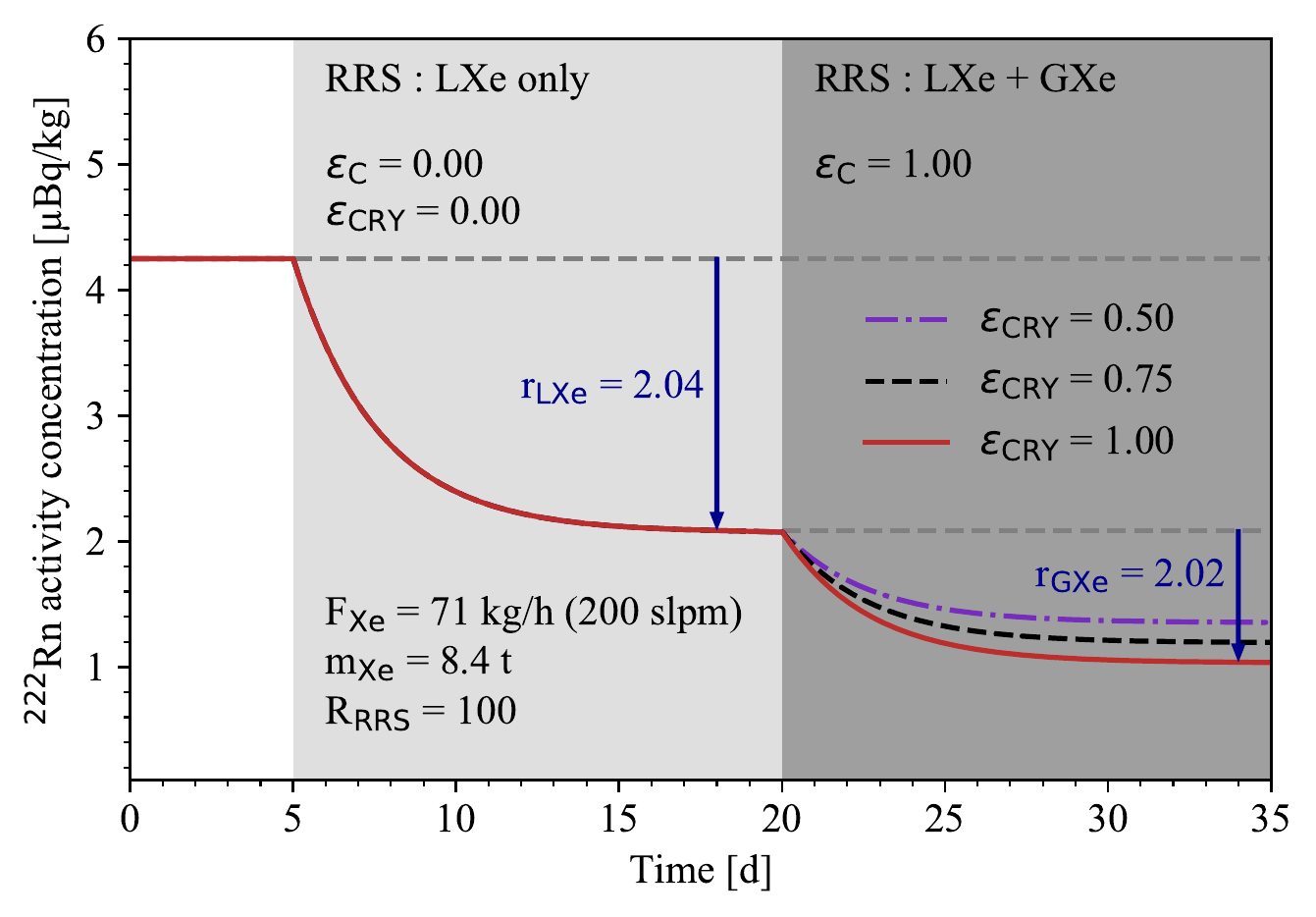}
 	\caption{Expected radon reduction in XENONnT for the LXe only mode as well as the LXe + GXe mode for $R_\mathrm{RSS}\gg 1$. For the LXe only mode, the design purification speed of the RRS of \SI{71}{kg/h} (200\,slpm) is assumed. For the LXe + GXe mode, a cable extraction efficiency of  $\epsilon_{\mathrm{C}}= 1$ is assumed. The total reduction is shown for extraction efficiencies of the CRY system of $\epsilon_{\mathrm{CRY}} = 0.5$ (purple), $\epsilon_{\mathrm{CRY}} = 0.75$ (black) and $\epsilon_{\mathrm{CRY}} = 1$ (red).}
 	\label{fig:expectedreduction}
\end{figure}

The expected radon reduction in XENONnT for a LXe only ($\epsilon = 0$) and a LXe + GXe ($\epsilon > 0$) mode is presented for different extraction efficiencies $\epsilon_{\mathrm{CRY}}$ and $\epsilon_{\mathrm{C}}$ in figure \ref{fig:expectedreduction} using units of \si{\upmu Bq/kg} instead of number of radon atoms.

In XENON1T, a full extraction from Cables was found ($\epsilon_{\mathrm{C}} = 1$), while the extraction from the CRY system $\epsilon_{\mathrm{CRY}}$ was about \num{0.5} \cite{mur, OnlRAD}. For XENONnT, the extraction flow from the CRY system is enhanced from \SI{0.5}{kg/h} (\SI{1.5}{slpm}) to \SI{7.1}{kg/h} (\SI{20}{slpm}) to get a more efficient extraction. For $\epsilon_{\mathrm{C}} = 1$ and $\epsilon_{\mathrm{CRY}}$ close to \num{1}, XENONnT can reach its projected \isotope[222]{Rn} activity concentration of \SI{1}{\upmu Bq/kg} with the help of the novel radon removal system.

\section{Modified McCabe-Thiele method without bottom product extraction}
\label{sec:McCabe}
The radon removal system for XENONnT is based on a cryogenic distillation column where the removal concept is based on trapping \isotope[222]{Rn} in the liquid xenon reservoir in the bottom of the system until it decays. This is possible due to its comparable short half-life of $T_{1/2}\,=\,\SI{3.8}{d}$. As consequence, there is no extraction of contaminant enriched xenon as it is required for krypton or argon distillation \cite{Pa, Xma, Kr1, Kr2}, and thus radon removal is a xenon-loss free operation. This requires a modification of the McCabe-Thiele method \cite{Mc}, a general approach to design such a distillation system. The modified approach is derived in this work.

In a single distillation stage with a static gas-liquid xenon interface at equilibrium, the size of the radon depletion in the gaseous xenon is described by the relative volatility $\alpha$ deduced from Raoult's law. It is the ratio between the vapor pressure of the noble gas contaminant radon $P_{\mathrm{Rn}}$ and the one of xenon $P_{\mathrm{Xe}}$:
\begin{align}
\alpha = \frac{P_{\mathrm{Rn}}}{P_{\mathrm{Xe}}} = 0.1 \ \ \ \text{at} \ \ \ \SI{-98}{\degreeCelsius}\text{\,\cite{NST}}.
\end{align}

Measurements were performed in a boil-off experiment with a single-stage setup \cite{br} yielding a radon depletion factor greater 4 in the gaseous phase, depending on the extraction flow, to verify the concept of relative volatility for radon concentrations on the order of \SI{e-25}{mol/mol} ($\mathcal{O}$($\upmu$Bq/kg)).

\begin{figure}[!h]
	\centering
    \includegraphics[width=\linewidth]{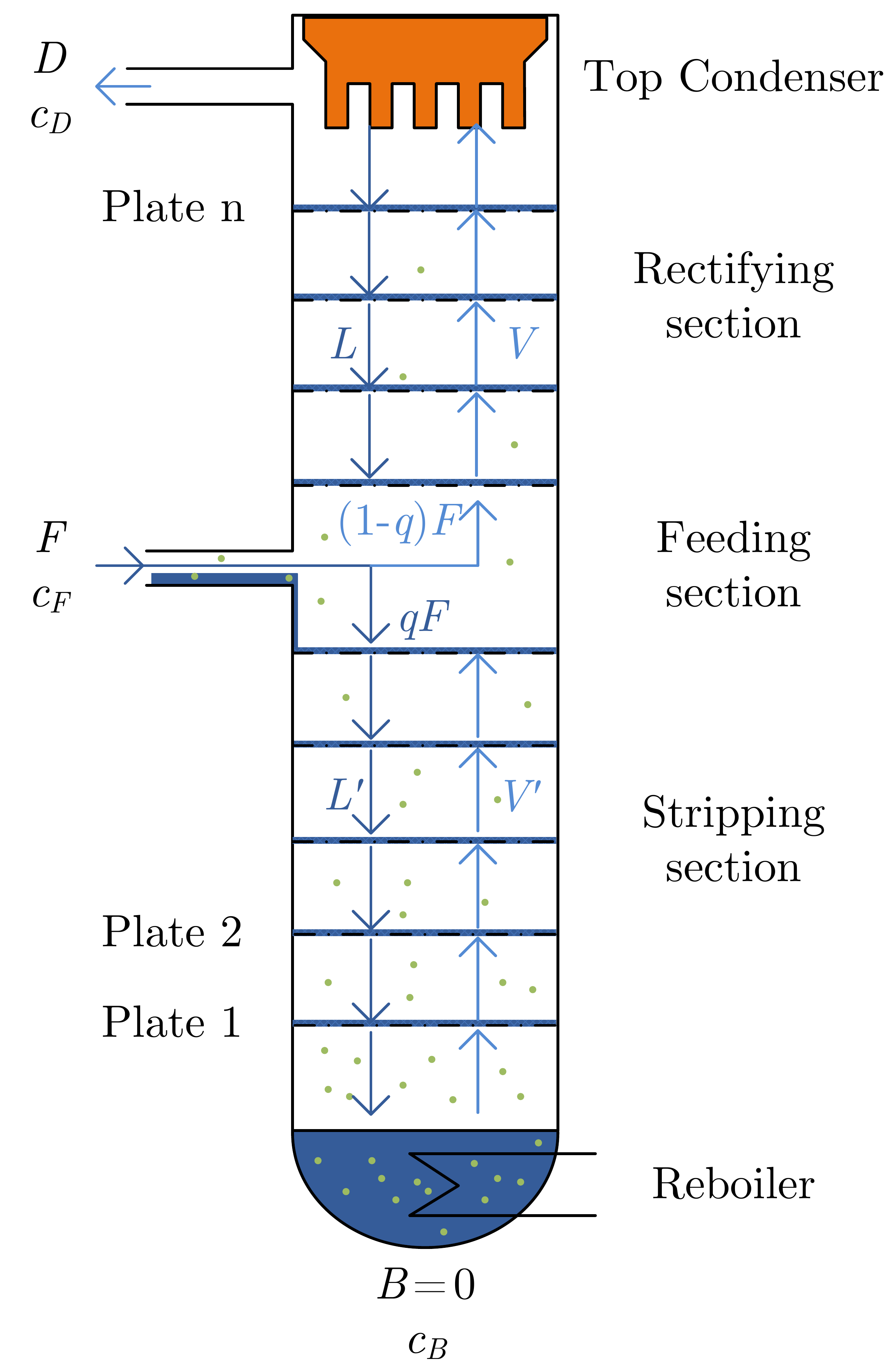}
 	\caption{Scheme of a multi-stage distillation tower with n plates, where the less volatile component is the contaminant (green dots). The incoming flow F with the contamination concentration $c_{\mathrm{F}}$ in the feeding section is enriched in the stripping section below to a contamination concentration $c_{\mathrm{B}}$ in the reboiler. In the case of radon, it is trapped inside the reboiler until it decays, and therefore, no bottom product extraction is required ($B=0$). In the rectifying section, above the feed, the contamination is depleted to a concentration $c_{\mathrm{D}}$ in the output flow $D$ at the top condenser. Furthermore, the liquid flows $L$, $L^\prime$ and gaseous flows $V$, $V^\prime$ of the related sections are shown. The factor $q$ indicates if the feed is gaseous ($q=0$) or liquid ($q=1$).}
 	\label{fig:column_scheme}
\end{figure}

In order to achieve larger separation factors, a distillation plant is made of a series of theoretical plates\footnote{One theoretical plate equals one distillation stage.} to successively reduce the impurities. The McCabe-Thiele method is applied to compute the number of required plates $n_{\mathrm{TP}}$ to achieve the desired concentration at the purified outlet. In this method, the column is separated in three parts, namely the feeding section, where the raw xenon gas is injected to the tower, the rectifying section, where the less volatile component, here the contaminant radon, is depleted and the stripping section, where the less volatile contaminant is enriched. This is depicted in figure \ref{fig:column_scheme} along with relevant xenon flows and radon concentrations for each section.

Starting from the reboiler stage in the stripping section at the bottom, the liquid xenon is evaporated and a gaseous stream flows upwards at a flow rate of $V^\prime$. Assuming that the vapor will condense on Plate 1, the generated liquid phase on this stage will have the same composition as the vapor from the reboiler stage, and will therefore have a lower concentration of the contaminant compared to the starting mixture due to the lower volatility. By repeating this procedure for several stages, the fraction of the contaminant in the xenon will successively decrease.

At the top, the concentration of the contaminant is at its minimum. While a part of the vapor is liquefied with the help of the top condenser and fed back to the column with a flow rate $L$, a fraction of the gas is removed as purified xenon at a flow rate $D$.
The ratio between the flows $L$ and $D$ is defined as the reflux ratio $R$. In the same manner, a second ratio $R^{\prime}$ can be found at the bottom. However, since no xenon is extracted at the bottom ($B=0$), the same flow that is entering the reboiler as liquid needs to be evaporated again. Therefore, $R^{\prime}$ becomes infinity:
\begin{equation}
R = \frac{L}{D} \qquad \textrm{and} \qquad R^{\prime} = \lim_{B \to 0} \frac{V^{\prime} }{B} = \infty.
\label{eq:reflux}
\end{equation}
The stripping section and the rectifying section are connected by the feeding section, where xenon is injected at a flow rate $F$. It is added to either the up-streaming vapour $V$ in the rectifying section or to the \linebreak down-streaming liquid flow $L^\prime$. This depends on the state of the injected matter (gaseous or liquid), denoted by the caloric factor
\begin{equation}
q = (L^\prime-L)/F.
\end{equation}

For the following calculations a saturated liquid feed at the boiling point $(q=1)$ is assumed. As consequence, the injected xenon is added to the down-streaming liquid flow $L^\prime$.
In the McCabe-Thiele diagram, the concentrations of the less volatile noble gas in the gaseous xenon phase ($y_{\mathrm{Rn}}$) and in the liquid xenon phase ($x_{\mathrm{Rn}}$) are plotted against each other in an equilibrium diagram.
For constant temperature and pressure, the system drifts to equilibrium based on the concept of vapor pressure as described above. For each local concentration $x_{\mathrm{Rn}} = (1-x_{\mathrm{Xe}})$ of the contaminant in the liquid phase, there is a certain equilibrium concentration $y_{\mathrm{Rn}}$ in the vapor phase. For very low concentrations with $x_{\mathrm{Rn}}\cdot(\alpha-1)\ll 1$ this relation can be expressed by 
\begin{equation}
    y_{\mathrm{Rn}} =\frac{\alpha\cdot x_{\mathrm{Rn}}}{1+(\alpha-1)\cdot x_{\mathrm{Rn}}} \approx \alpha\cdot x_{\mathrm{Rn}}.
\label{eq:equilibriumline}
\end{equation}
In the diagram, this is called the equilibrium line.

Due to active heating from the bottom and cooling from the top, the upward gaseous ($V, V^\prime$) and downward liquid ($L, L^\prime$) flows are created along the column and as consequence, the equilibrium inside the system is disturbed due to the mass flow. As a result of that, the actual concentration of the contaminant in the gaseous phase related to the concentration in the liquid phase is described by operation lines for each of the three sections. Between the equilibrium line and operation lines a forcing concentration gradient is ensured, driving the system toward the equilibrium via mass transfer between the ascended gaseous stream and the descended liquid stream.

During stable operation, the incoming xenon feed flow $F$ requires to be equal to the outgoing xenon flow $D$ yielding the total mass balance
\begin{equation}
F = D.
 \label{eq:radon_mass_balance}
\end{equation} 

Combining the flows with the different concentrations of the less volatile noble gas radon inside the xenon in the feed $c_F$, in the top $c_D$ and in the bottom $c_B$, equation (\ref{eq:radon_mass_balance}) can be expressed in terms of the number of contaminant atoms. In this work, the radon particle decay inside the reboiler needs to be taken into account to ensure particle mass balance:
\begin{equation}
 F\cdot c_F=D\cdot c_D + c_B \cdot \lambda_{\mathrm{Rn}} \cdot m_{\mathrm{Xe,reb}}.
 \label{eq:radon_particle_balance}
\end{equation}
Here, $\lambda_{\mathrm{Rn}}\,=\,\SI{7.55E-03}{\per\hour}$ denotes the decay constant of \isotope[222]{Rn} and $m_{\mathrm{Xe,reb}}$ is the liquid xenon mass inside the reboiler. The last term effectively describes an out-going mass flow of radon particles from the system by decay and is the equivalent to an out-going offgas mass flow in the standard McCabe-Thiele approach, e.g. for krypton \cite{mur}. The decay of radon particles in the other parts of the distillation system can be neglected due to the short residence time compared to the decay time.

When fixing $F$, $c_F$, and $c_D$ by design, the mass of radon particles in the reboiler $c_B \cdot m_{\mathrm{Xe,reb}}$ requires to be calculated with above equations in order to conserve the particle mass balance:
\begin{equation}
c_B \cdot m_{\mathrm{Xe,reb}} = \frac{c_F \cdot F - c_D \cdot D}{\lambda_{\mathrm{Rn}}}.
\label{eq:radon_concentration_reboiler}
\end{equation}
For the calculation of the McCabe-Thiele diagram, either the concentration $c_B$ or the mass in the reboiler $m_{\mathrm{Xe,reb}}$ has to be fixed by design.

For each of the three sections in the system, a particle balance can be formulated using the respective flows in the sections. From these, the three operation lines can be obtained:
\begin{align}
\label{eq:strpping_line}
\textrm{Stripping line:} \, \, y_{\mathrm{Rn}} &= x_{\mathrm{Rn}} - \frac{c_B \cdot \lambda_{\mathrm{Rn}} \cdot m_{\mathrm{Xe,reb}}}{L^{\prime}},\\
\label{eq:rectifying_line}
\textrm{Rectifying line:}  \, \, y_{\mathrm{Rn}} &= \frac{R}{R+1} \cdot x_{\mathrm{Rn}} +  \frac{c_D}{R+1},\\
\label{eq:feed_line}
\textrm{Feeding line:}  \, \, y_{\mathrm{Rn}} &= \frac{q}{q-1} \cdot x_{\mathrm{Rn}} - \frac{c_F}{q-1}.
\end{align}
More details for the derivation of the operation lines can be found in Ref. \cite{mur}.

In the following, the  McCabe-Thiele diagram for the RRS for XENONnT is discussed. The system was designed for a feed flow of $F\,=\,\SI{71}{kg/h}$ (200\,slpm). 
As described in section \ref{sec:radonremovalstrategy}, an initial activity concentration of $ c^{\prime}_F = \SI{4.25}{\upmu Bq/kg}$ is expected. This can be converted into a radon-in-xenon concentration of $c_F=\SI{4.4e-25}{mol/mol}$ via 
\begin{equation}
    c_F = \frac{M_\mathrm{Xe}}{ \lambda_{\mathrm{Rn}} \cdot N_\mathrm{A}} \cdot c^{\prime}_F = \SI{1.04e-25}{\frac{mol}{mol}\frac{kg}{\upmu Bq}} \cdot c^{\prime}_F,
\end{equation}
with $M_\mathrm{Xe} = \SI{131.3}{g/mol}$ being the xenon molar mass, $\lambda_{\mathrm{Rn}} = \SI{2.098e-6}{1/s}$ being the \isotope[222]{Rn} decay constant, and $N_\mathrm{A}$ being the Avogadro number.

In order to meet at least the requirements for a factor two reduction as derived in section \ref{sec:radonremovalstrategy}, a depletion of a factor \num{100} at the top and an enrichment of a factor \num{1000} at the bottom with respect to the feed concentration were chosen. Since the removal is based solely on the \isotope[222]{Rn} decay, the RRS requires to have a large enrichment factor between inlet and bottom part. This allows to reduce the required xenon inventory inside the RRS system while avoiding back contamination from the highly enriched bottom.
It follows $c_D = \SI{4.4e-27}{mol/mol}$ and $c_B = \SI{4.4e-22}{mol/mol}$. \linebreak The reflux ratio $R$ determines the amount of xenon that needs to be re-liquefied and send back to the column and is therefore directly proportional to the required cooling power at the top condenser. For high purification flows and large $R$, this cooling can start to become an issue. Therefore, a minimal reflux ratio $R_{\mathrm{min}}$ has to be estimated for the radon column design to minimize the needed cooling power. The Underwood equation \cite{Mc} can be utilized to calculate the reflux ratio that results in the operation line with minimal slope:
\begin{equation}
R_{\mathrm{min}} = \frac{1}{\alpha - 1} \cdot \left(\frac{c_D}{c_F} - \alpha \cdot \left(\frac{1 - c_D}{1 - c_F} \right)\right) = 0.1.
\end{equation}

Several McCabe-Thiele diagrams for a liquid feed ($q=1$) and the given concentrations above were constructed with varying reflux ratios starting with  $R=0.1$ up to $R=5$. For $R<0.1$ no diagrams can be constructed since the rectifying line starts to go below the equilibrium line and no stages can be drawn anymore. For $R=0.11$ a total number of 16 stages is required to meet the desired purity level. The number of stages decreases with increasing $R$ up to $R=0.5$. Here, 7 theoretical stages can be drawn. With further increase of the reflux, the number of stages can be reduced to 6 for $R=5.0$. The reduction by one stage between $R=0.5$ and $R=5.0$ would require a ten times higher cooling power at the top condenser. Therefore, a reflux ratio of $R=0.5$ was chosen for the design also allowing for larger feed flows in the future. All other required parameters to create the diagram for the chosen reflux ratio are summarized in table \ref{tab:diagram}.

\begin{table}[htbp]
\centering
\caption{Input parameters for McCabe-Thiele diagram of the XENONnT RRS.}
\begin{tabular*}{\columnwidth}{ll}
\hline
 Section    & Parameter \\ \hline
 Feeding    & $c_F= \SI{4.4e-25}{mol/mol}$ \\ 
            & $F= \SI{71}{kg/h} \, (\SI{200}{slpm})$ \\
            & $q = 1$ \\ \hline
 Stripping  & $c_B= \SI{4.4e-22}{mol/mol}$ \\ 
            & $B = \SI{0}{kg/h} $\\
            & $L^{\prime} = \SI{106}{kg/h}  \, (\SI{300}{slpm})$ \\
            & $V^{\prime} = \SI{106}{kg/h} \, (\SI{300}{slpm})$ \\
            & $R^{\prime}=\infty$ \\
            & $m_{\mathrm{Xe,reb}} = \SI{9.4}{kg}$ \\ \hline
 Rectifying & $c_D=\SI{4.4e-27}{mol/mol}$ \\ 
            & $D = \SI{71}{kg/h}  \, (\SI{200}{slpm})$ \\
            & $L = \SI{35}{kg/h}  \, (\SI{100}{slpm})$ \\
            & $V = \SI{106}{kg/h}  \, (\SI{300}{slpm})$\\
            & $R=0.5$ \\ \hline
\end{tabular*}
\label{tab:diagram}
\end{table}

\begin{figure}[!h]
	\centering
    \includegraphics[width=\linewidth]{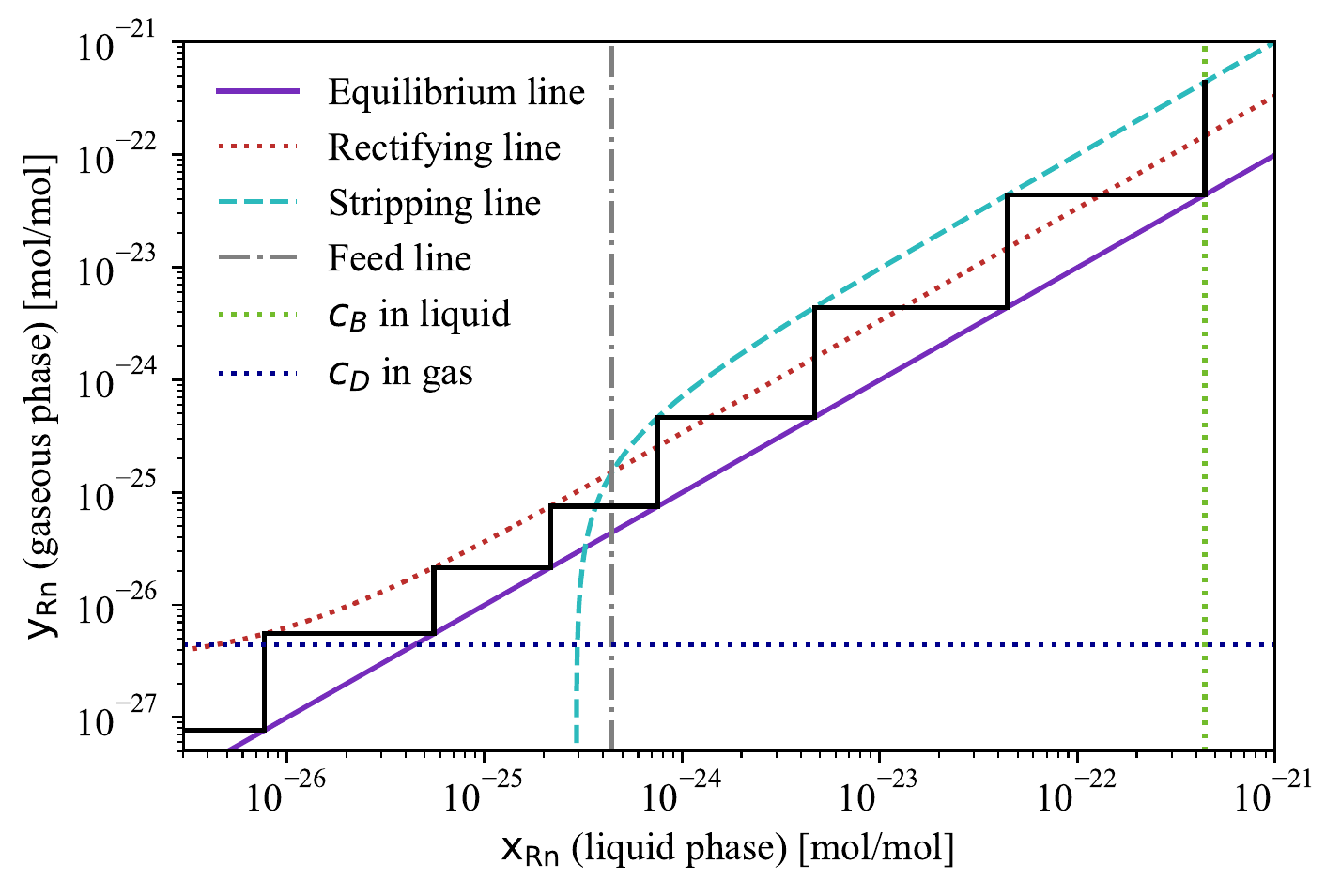}
 	\caption{ McCabe-Thiele diagram for a radon distillation system with R\,=\,0.5: The equilibrium line (solid purple) is below the stripping line (dashed light blue), and the rectifying line (dotted red). The operation lines are intersected by the feed line (dash dotted gray), which is a vertical line due to $q = 1$. One vertical line (dotted green) represents the concentration $c_B$, while the horizontal line (dotted dark blue) gives the concentration $c_D$. The bottom of the column corresponds to the right and the top of the system to the left part of the diagram.}
 	\label{fig:McCabe_Thiele_rn}
\end{figure}

The resulting McCabe-Thiele diagram is visualized in figure \ref{fig:McCabe_Thiele_rn}. The solid purple equilibrium line in contrast to the case of a more volatile component like krypton is below the operation lines due to $\alpha\,=\,\num{0.1}$. The stripping line from the bottom part, where in this case radon is enriched, is indicated by the light blue dashed line, while the rectifying line from the top, where radon is depleted, is drawn as red dotted line. The operation lines are intersected by the gray dash dotted feed line, which is a vertical line due to $q\,=\,1$. In contrast to krypton, the concentration $c_B$ in the liquid, visualized for better understanding by the green dotted vertical line, is located at the right border of the diagram. The concentration $c_D$ in the gas, drawn as dark blue dotted horizontal line, is located at the left border. Consequently, the bottom of the distillation column corresponds to the right and the top to the left part of the diagram. In total, 7 theoretical stages can be drawn, where 4 stages below and 3 stages above the feed section are required to achieve the desired enrichment and depletion factors of \num{1000} and \num{100}, respectively.

From the experimental point of view, a series of multiple physically connected single distillation stages is not practical. For this reason, a structured packing material inside a package tube is utilized, featuring a large surface for good liquid-gas exchange through the full height of the tower. The height-equivalent of one theoretical plate (HETP) denotes the amount of required packing material to realize one distillation stage. Together with the number of plates $n_{\mathrm{TP}}$ derived from the McCabe-Thiele diagram, the total height $h$ of the package tube can be calculated:
\begin{equation}
 h=\mathrm{HETP}\cdot n_{\mathrm{TP}}.
 \label{eq:HETP}
\end{equation}
Here, one stage is intrinsically given by the reboiler combined with the top condenser. Thus, six theoretical plates have to be considered for the package tube. The HETP value of the used packing material is unknown. However, tests at XENON100 with a distillation column for krypton in inverse mode \cite{OnlRAD} give an upper limit for the HETP value of \SI{27.5}{cm} \cite{mur}. A similar packing material from the same company, but with larger diameter is used for the radon distillation plant of XENONnT. Taking the upper limit, a maximum height of \SI{165}{cm} is calculated for 6 stages for the new package tube.

\section{The Radon Removal System}
\label{sec:radon_column}
The cooling concept as well as the experimental realization of the RRS is described in this section. The design parameters used for the McCabe-Thiele diagram in section \ref{sec:McCabe} define the requirements in terms of flows, cooling and heating power. In order to reach the large target flow of $F = \SI{71}{kg/h}$ (\SI{200}{slpm}), the RRS is operated with LXe inlet and outlet. This is possible due to the novel LXe purification system in XENONnT, where a LXe fraction is diverted into the RRS. The returning radon-depleted LXe is then mixed back into the LXe circuit before entering the cryostat. The total required cooling and heating powers can be derived from the involved flows in the design.

At the top condenser, a cooling power of about \SI{1}{kW} is required to guarantee the reflux ratio of $R=0.5$ resulting in a liquefaction flow of $L = \SI{35}{kg/h}$ (\SI{100}{slpm}) \cite{k}. The radon-depleted GXe xenon flow of $D = \SI{71}{kg/h}$ (\SI{200}{slpm}) extracted at the top condenser has to be liquefied again before it can be returned to the LXe purification circuit. This requires an additional cooling power of about \SI{2}{kW}, leading to a total required cooling power of \SI{3}{kW}. Additionally, a heating power of about \SI{3}{kW} at the reboiler is required to create the evaporation flow $V^{\prime} = \SI{106}{kg/h}$ (\SI{300}{slpm}) to establish a stable distillation process across the package tube.

For the cooling, either an industrial cooling machine or liquid nitrogen (LN$_{2}$) can be used. However, industrial solutions were not readily available for our application and would have to be customized for our purposes. Furthermore, often the customer does not have full control over the components used and the manufacturing processes applied by the company. The selection of high purity material and manufacturing following standards of ultra-high vacuum components are key requirements for the RRS. Therefore, it was decided to use LN$_{2}$ and to develop a custom-made cooling solution. Furthermore, LN$_{2}$ provides a simple cooling mechanism, is cost-efficient and has low-maintenance requirements in the long term perspective compared to industrial cooling machines. It further gives the possibility to scale up in terms of supplied cooling power at a later stage. The XENONnT infrastructure features a 30 tonne LN$_{2}$ tank underground already, serving several other subsystems. To reach \SI{3}{kW} of cooling, a LN$_{2}$ consumption of \SI{1300}{kg/d} is required taking only the LN$_{2}$ phase change into account with an enthalpy of evaporation $\Delta H_\mathrm{vap}^{\mathrm{N}_2}$ of \SI{199}{kJ/kg} at \SI{77}{K} for LN$_{2}$.

\subsection*{Cooling Concept}
\begin{figure*}[!h]
	\centering
	\includegraphics[width=\linewidth]{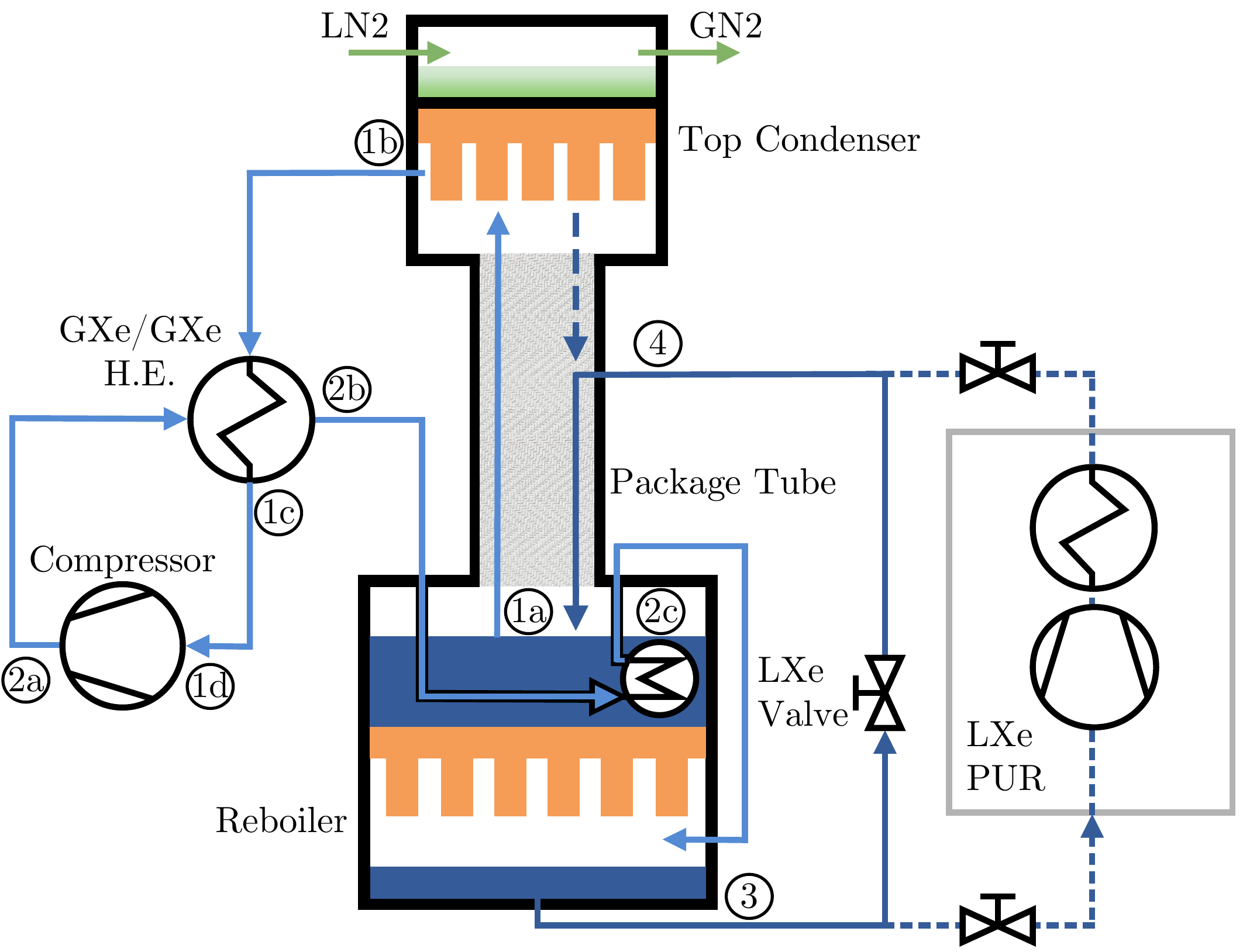}
 	\caption{Layout of the radon distillation system. The four main components are the top condenser, the reboiler, the package tube and the compressor. The cooling concept is based on a Clausius-Rankine cooling cycle with phase-changing cooling medium indicated by the numbers (1) to (4).}
 	\label{fig:coolingconcept}
\end{figure*}

An energy-efficient cooling concept for the RRS was designed, as shown in figure\,\ref{fig:coolingconcept}, to drastically reduce the LN$_{2}$ consumption. In this concept, the reboiler acts as a bath-type Xe-Xe heat exchanger (HE) with a top and bottom vessel, both thermally connected with copper:  Cooling energy stored in the radon-enriched LXe inside the top reboiler vessel is used to provide \SI{2}{kW} of cooling power for the radon-depleted GXe liquefaction process inside the bottom reboiler vessel. Like that, in parallel, the LXe in the top reboiler vessel is evaporated and the required electrical power to establish the evaporation flow $V^{\prime}$ is reduced to \SI{1}{kW} heating power. Making use of this cooling concept reduces the LN$_{2}$ consumption to \SI{430}{kg/d} applied at the top condenser to create the desired reflux. However, in order to make the reboiler HE work, a temperature difference between the two xenon phases needs to be established such that heat can be transferred from the bottom to the top reboiler vessel. Therefore, the concept further utilizes a heat-pump, or compressor, that extracts radon-depleted GXe from the top condenser and compresses it into the reboiler bottom vessel. Furthermore, two commercial GXe-GXe HE are applied to bring the extracted GXe for the compression process to room temperature and to pre-cool the returning compressed GXe. A spiral is immersed in the LXe of the top reboiler vessel to even further cool the compressed GXe before it enters the bottom reboiler vessel. After liquefaction in the bottom reboiler vessel, the radon depleted LXe is returned to the LXe purification circuit. A LXe bypass valve was installed to connect the RRS outlet directly with the inlet to allow for performance tests independent from the global XENONnT system.

The thermodynamic concept can be described by a left turning Clausius-Rankine cycle with a phase-changing cooling medium, in this case the xenon itself. Beside the four usual steps of evaporation (\textcircled{\raisebox{-0.8pt}{4}}$\rightarrow$\textcircled{\raisebox{-0.8pt}{1}}), compression (\textcircled{\raisebox{-0.8pt}{1}}$\rightarrow$\textcircled{\raisebox{-0.8pt}{2}}), condensation (\textcircled{\raisebox{-0.8pt}{2}}$\rightarrow$\textcircled{\raisebox{-0.8pt}{3}}) and expansion (\textcircled{\raisebox{-0.8pt}{3}}$\rightarrow$\textcircled{\raisebox{-0.8pt}{4}}), the thermodynamic cycle of the RRS illustrated in figure \ref{fig:coolingconcept} features multiple intermediate states of heat transfer making the cycle more complex. The different states are summarized in table \ref{tab:states_summary}. A p-H- and a T-S-diagram to describe the cycle are presented in figure \ref{fig:ph_ts} of section \ref{sec:commissioning} using data from the RRS's performance tests.

\begin{table}[htbp]
\caption{States of the Clausius-Rankine cycle for the RRS cooling concept. }
\centering
\begin{tabular*}{\columnwidth}{lll}
\hline
 State &  & Description\\ \hline
  \textcircled{\raisebox{-0.3pt}{\scriptsize{1a}}} & evaporation &  LXe evaporation in top reboiler\\ 
  \textcircled{\raisebox{-0.3pt}{\scriptsize{1b}}} &    & GXe extraction \\ 
 \textcircled{\raisebox{-0.3pt}{\scriptsize{1c}}} &    & GXe HE warm-up\\ 
 \textcircled{\raisebox{-0.3pt}{\scriptsize{1d}}} &    & 4MP inlet\\ \hline
 \textcircled{\raisebox{-0.3pt}{\scriptsize{2a}}} & compression & 4MP outlet\\ 
 \textcircled{\raisebox{-0.3pt}{\scriptsize{2b}}} &  & GXe HE pre-cooling\\ 
 \textcircled{\raisebox{-0.3pt}{\scriptsize{2c}}} & & GXe spiral pre-cooling\\ \hline
 \textcircled{\raisebox{-0.3pt}{\scriptsize{3}}} & condensation & Liquefaction in bottom reboiler\\ \hline
 \textcircled{\raisebox{-0.3pt}{\scriptsize{4}}} & expansion & Internal bypass\\ \hline
\end{tabular*}
\label{tab:states_summary}
\end{table}

\subsection*{Experimental Setup}
\begin{figure*}[!h]
	\centering
	\includegraphics[width=\linewidth]{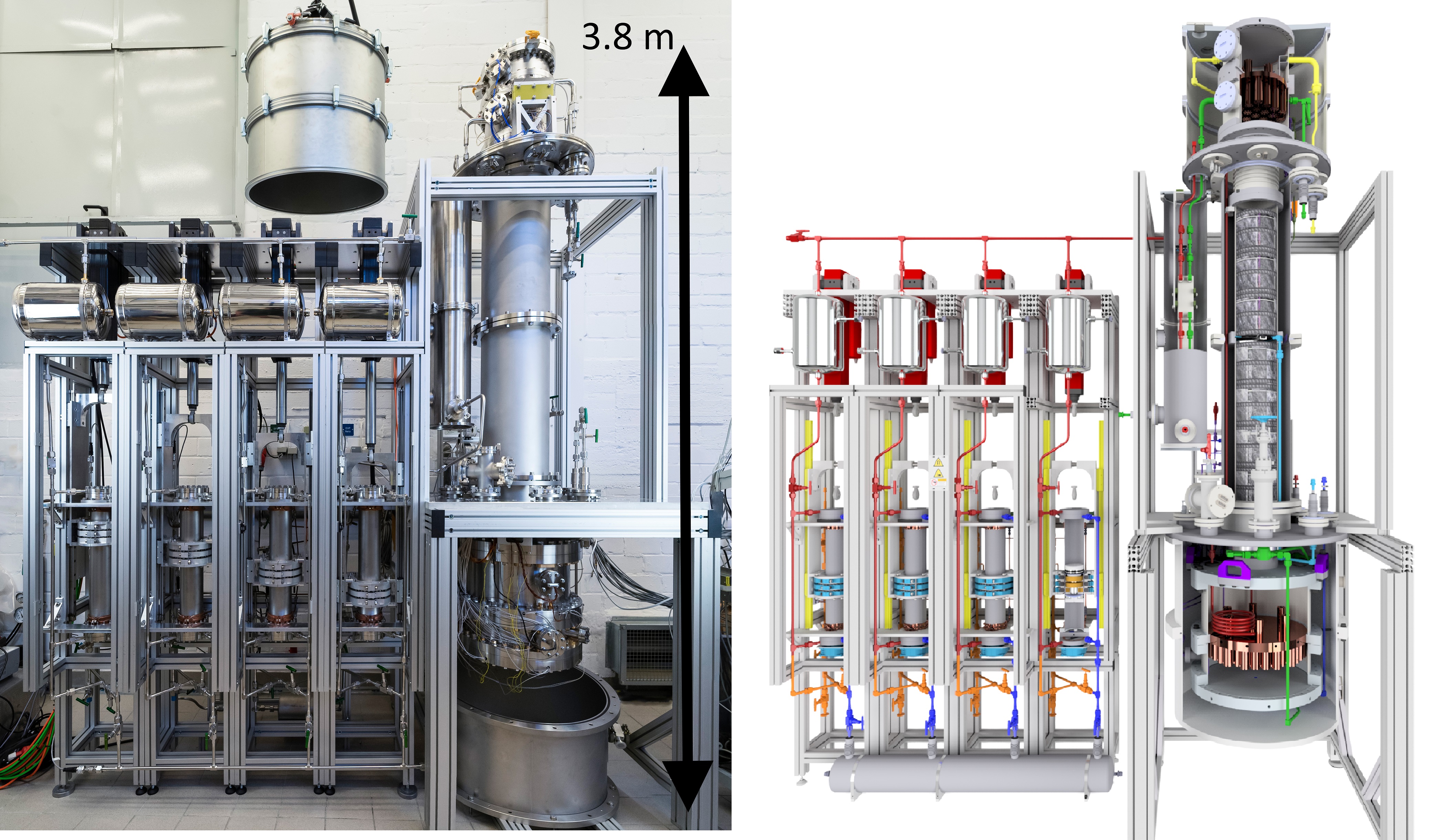}
 	\caption{Radon removal system fully installed at University of Münster before shipping to LNGS (left) and related CAD drawing (right). The four key components top condenser, package tube, reboiler and four-cylinder magnetically-coupled piston pump are visible.}
 	\label{fig:radfull}
\end{figure*}

The four main components of the RRS for XENONnT are the top condenser, the reboiler, the package tube and the compressor. Additionally, auxiliary components are used to warm up and pre-cool the GXe in the cycle. The experimental setup is described in the following. The full radon removal system was first constructed at the University of Münster as shown in figure\,\ref{fig:radfull}. Later, it was partially disassembled, shipped and installed inside the top floor of the XENONnT service building. The system's total height is \SI{3.8}{m} constrained by the floor height inside top floor.
In order to monitor the RRS during operation, the entire system is equipped with 32 Resistance Temperature Detectors (RTDs) (Honeywell HEL-705), four silicon diodes (Lakeshore DT670D-CU), 13 pressure sensors (Wika WU20), four differential pressure sensors (Baratron 226A), and a vacuum pressure sensor (Thyracont VSM77DL), all controlled by a Programmable Automation Controller (PAC) from General Electric (GE RX3i) on the hardware side and Cimplicity SCADA (Supervisory Control And Data Acquisition) on the software side in the private front-end network of XENONnT.

Furthermore, to minimize external heat inputs, the inlet and outlet LXe tubes connecting the RRS to the LXe Purification system as well as the full distillation system are encapsulated in an insulation vacuum. The top condenser insulation chamber features a diameter of 650\,mm and a height of 658\,mm, the reboiler vessel a diameter of 750\,mm and a height of 758\,mm and the two-component package tube vessels a diameter of 300\,mm and a height of 712\,mm each.
\subsubsection*{Package Tube}

The package tube extending into the top condenser insulation vessel as well as into the reboiler vacuum chamber has a total height of \SI{1900}{mm} and is made of three segments: The bottom segment that connects to the reboiler is a stainless steel tube with a height of 920\,mm and diameter of \SI{200}{mm}. A custom-modified Conflat (CF) flange at its top features two feedthroughs for GXe and LXe feed tubes. In order to spread the incoming LXe around the whole package diameter equally, a distributor is connected to the LXe feedthrough on the inside. 
The top segment is a stainless steel bellow with eight windings, a wall thickness of 0.2\,mm and a nominal operation pressure of 4\,bar that connects to the top condenser. It is used to compensate thermal shrinkage and expansion when cooling down the system from room temperature to liquid xenon temperature or vice versa. Eight package segments (Sulzer PDX) of \SI{178}{mm} height and \SI{200}{mm} diameter each were inserted into the two stainless steel tubes, five in the bottom tube and three in the top tube, leading to a total package height of \SI{1420}{mm}. 
Furthermore, liquid xenon pneumatic flow controlled valves (Thermomess) at the liquid inlet and outlet are used to regulate the flow-rate through the system during operation within the global system.

\subsubsection*{Top Condenser}
The top condenser described in detail in Ref. \cite{k} with a diameter of \SI{300}{mm} is composed of two vessels: A cooling vessel using liquid nitrogen on top and a xenon liquefaction vessel below. Aluminum spacers keep the distance between the heat exchanger and the surrounding insulation vessel. In order to reduce the heat transfer via conduction between the inner vessels and the insulation chamber, these spacers are equipped with preimpregnated fibre plates. Additionally, the vacuum-insulated gas lines of the inlet and outlet for the xenon as well as the nitrogen are guided through the insulation chamber via bellow-based feedthroughs. 

For the cooling vessel at the top, the nitrogen supply line is given by a 1/2" tube, the exhaust line by a 1" tube. A total of \SI{7.1}{l} LN$_2$ can be filled. The vessel inside is equipped with baffle plates at the inlet and at the outlet preventing the nitrogen to bypass the system. Additionally, to increase the heat transfer surface, six oxygen-free high conductivity copper fins are mounted on the ground of the vessel.

The ground flange of the vessel, acting as the thermal connection to the liquefaction vessel, is given by a stainless steel blank flange of \SI{27}{mm} thickness with helicoflex-type sealing faces on both sides. Here, advanced gaskets (HTMS CSI) based on a copper-coated helical spring  with the capability to recover from thermal deformations are used.

Attached to the thermal connection flange a copper plate of \SI{290}{mm} diameter and \SI{20}{mm} thickness extends into the \SI{220}{mm} high xenon liquefaction vessel. Additional \num{23} copper fins mounted to the copper plate increase the heat transfer surface in contact to the xenon to \SI{0.76}{m^2}.

By regulating the LN$_2$ flow through the top vessel, variable cooling powers can be achieved. Cooling power measurements in Ref. \cite{k} show a performance of \SI{3}{kW} and beyond only limited by the available heating power to balance the system. This is three times larger than the required \SI{1}{kW} and gives the possibility to scale up the process flow in the future. During these performance tests, a heat exchanger efficiency of ($0.98\pm 0.03$) was computed. 

\subsubsection*{Reboiler}
Also the reboiler described in detail in Ref. \cite{k} with a diameter of \SI{500}{mm} is composed of two vessels: The storage vessel of the highly radon-enriched xenon on top and the xenon liquefaction vessel below at the bottom.

The top storage vessel is open to the package tube via a CF200 flange. A total of \SI{46.5}{l} LXe can be stored in the 237\,mm high top vessel. Furthermore, six oxygen-free high conductivity copper fins increasing the heat transfer surface in contact with the stored liquid are installed on the top side of the connecting flange. 

This connecting flange thermally connects and hermetically seals both reservoirs. It is made of a \SI{35}{mm} thick copper plate of \SI{450}{mm} diameter that was electron-welded into the connecting flange to improve the heat transfer. In total, \num{79} copper fins are attached to the down-facing side of the copper plate extending into the \SI{200}{mm} high liquefaction vessel in the bottom. This increases the total condensation surface in the bottom vessel in contact with the xenon to \SI{2.2}{m^2}. The liquefaction reservoir can be filled with up to \SI{19.6}{l} LXe limited by the length of the copper fins. The bottom flange is funnel-shaped to collect the radon-depleted LXe at a centered feed returning the liquid via a 1/2" stainless steel tube to the global system.

Furthermore, a PID regulated heating system based on 12 1/2" heating cartridges (WEMA S5105) connected to a programmable 3\,kW power supply (Kniel VE3PUI2) is used to balance the desired evaporation in the top reservoir as well as the liquefaction in the bottom reservoir depending on the circulation flow, the operation pressure and the applied cooling power at the top condenser.

Following the heat transfer model in Ref. \cite{k} based on conservative assumptions, an expected heat transfer between the two reservoirs of \SI{3293}{W} is expected for a pressures of \SI{2}{bar} in the top and \SI{3.5}{bar} in the bottom vessel. The reboiler performance can only be tested with the RRS system fully constructed. The results are presented in section \ref{sec:commissioning}.

\subsubsection*{Compressor}
The compressor is a four-cylinder magnetically-coupled piston pump based on a prototype used for the \linebreak XENON1T experiment \cite{BRW}. It features particular cleanliness and radiopurity to avoid contamination of the purified xenon. Further improvements of its dead volume and flow path compared to the prototype in combination with the usage of four pumps in parallel with a smaller dimension allows to reach the required flows and compression for the RRS design. A synchronized movement with a phase shift of the four pumps results in an additional performance boost and guarantees a stable operation of the RRS. The system features a flow meter (Brooks SLA5863) at its outlet, where flows of up to \SI{167(3)}{kg/h} (\SI{474(9)}{slpm}) were measured during commissioning at a xenon inlet pressure of \SI{2.1}{bar} and a compression of \SI{1.80(6)}{bar}. A detailed description of the design and performance is given in Ref. \cite{l}.

\subsubsection*{Auxiliary Components}
Two commercial plate heat exchangers (EWT BE4-10x20) featuring 1/2" inlets and outlets are installed in parallel between the column and the compressor to warm-up and pre-cool the GXe.

The spiral immersed in the top reboiler vessel is a 1/2" stainless steel tube with 4.25 windings and a total length of \SI{3000}{mm}. 

The LXe bypass valve (Thermomess) is specifically made for LXe applications and has a diameter of 6\,mm.

\section{Thermodynamic Performance}
\label{sec:commissioning}
The top condenser and the compressor were tested stand-alone as described in Ref. \cite{k} and Ref. \cite{l}, respectively. The performance of the reboiler can only be investigated with the full RRS system installed underground at LNGS. The thermodynamic functionality and stability of the full RRS is presented in this section. For this, the internal LXe bypass valve is used as the throttle valve, expanding and feeding the LXe from the bottom reboiler vessel at a high pressure to the LXe feed port of the package tube at a lower pressure. To start the LXe circulation through the bypass valve, the RRS needs to be cooled down, and both reboiler vessels have to be filled with LXe initially.

\subsection*{Filling}
In order to cool down the RRS from room temperature to about \SI{178}{K}, the entire system is filled with \SI{3}{bar} xenon. Then, the LN$_2$ cooling at the top condenser is activated, and the xenon starts to cool down. In the cooling process, LXe droplets continuously form at the top and fall down into the package tube, where they evaporate again. Like that, the top condenser, package tube and top reboiler vessel can be cooled down from top to bottom within \SI{12}{h}. Once the system is cold, GXe is continuously injected into the package tube via the GXe feed port, is then liquefied at the top condenser, and from there, rinsing down along the package tube into the top reboiler vessel.  

\begin{figure*}[t]%
    \centering
    \subfloat[\centering Filling]{{	\includegraphics[width=0.47\textwidth]{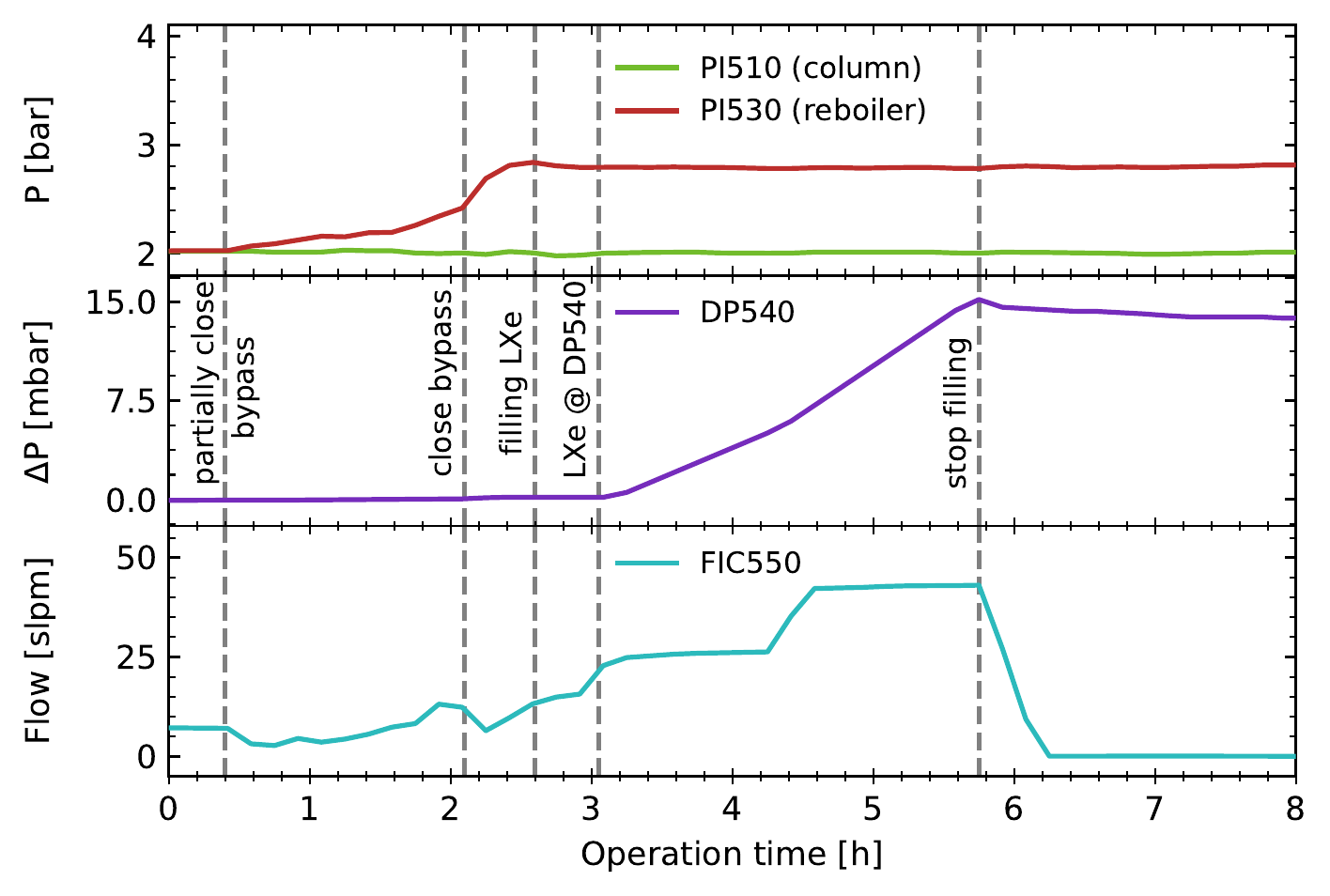} }}%
    \qquad
    \subfloat[\centering Internal liquid circulation]{{ \includegraphics[width=0.47\textwidth]{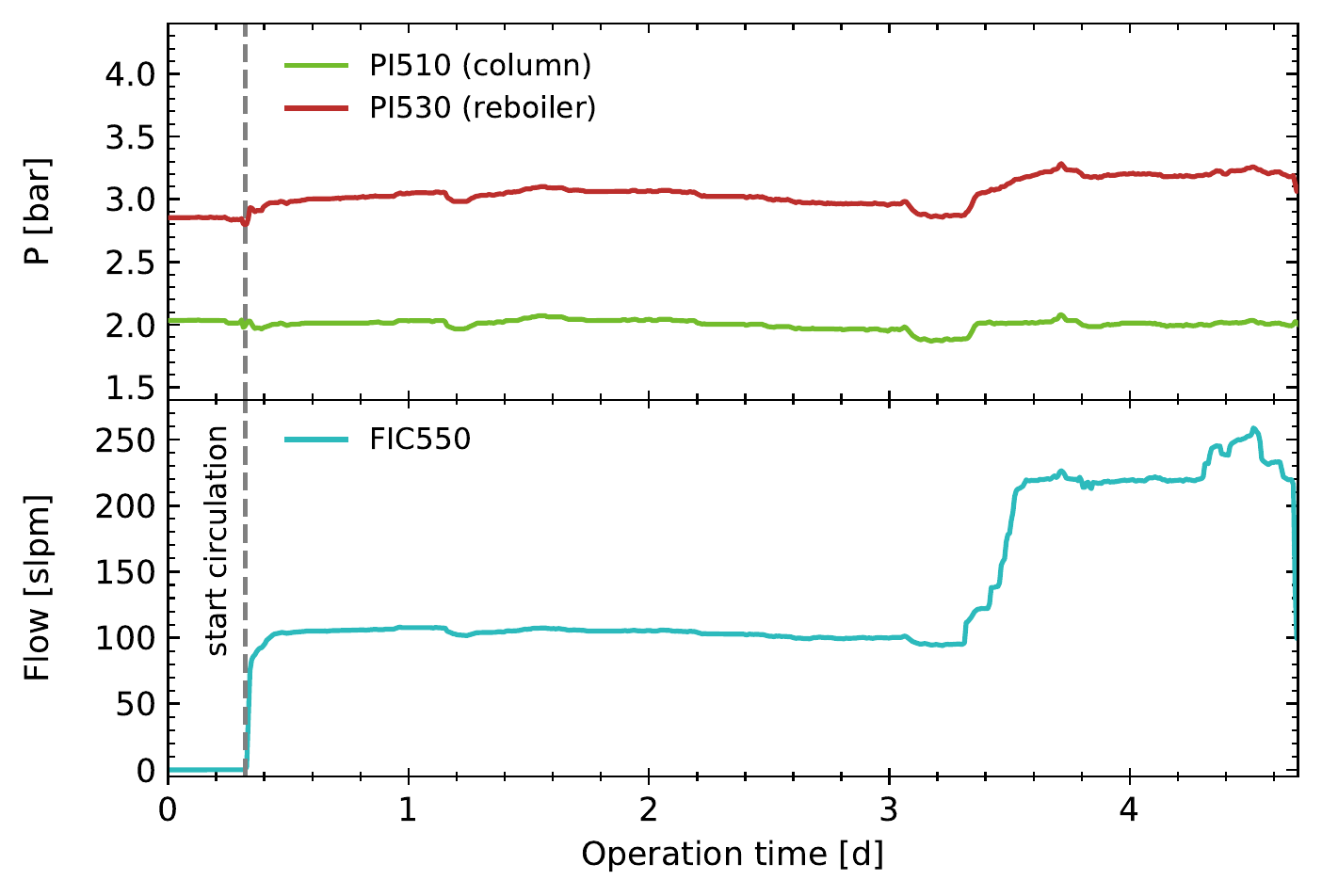}}}%
    \caption{RRS performance measurements: (a) Filling of the bottom reboiler vessel. By closing the LXe bypass valve, the bottom reboiler pressure (PI530)  increases up to \SI{2.84}{bar} required to liquefy the xenon gas flow (FIC550). The liquid level increase in the bottom part is indicated by a differential pressure measurement (DP540). (b) Start of internal liquid circulation of the RRS through the LXe bypass valve. The required bottom reboiler pressure (PI530) for the liquefaction of the required flow as well as the column pressure (PI510) are shown (top). Several circulation flows (FIC550) were tested (bottom). }%
    \label{fig:fill_circ}%
\end{figure*}

In the next step, xenon needs to be transferred into the bottom reboiler vessel, to run the internal circulation. This bottom filling is the first test to liquefy xenon with xenon in the reboiler system and is presented in figure\,\ref{fig:fill_circ}(a): The top panel contains the relevant pressures for the filling process, namely the column pressure PI510 (green), and the bottom reboiler pressure PI530 (red). The middle panel shows the liquid level meter measurements DP540 (purple) based on differential pressure in the bottom reboiler vessel. The bottom panel illustrates the flow FIC550 (light blue) created by the compressor. Before a continous filling can be established, the bottom reboiler vessel is pre-cooled as well by circulating GXe through the entire system. In detail, cold GXe is extracted and warmed up using the GXe/GXe HE. The compressor pushes the GXe back into the GXe/GXe HE where it is cooled down again. The GXe enters the spiral to be even further cooled, and then enters the bottom reboiler. From here, the GXe flows back into the package tube via the fully opened bypass valve. This is done over the course of a few hours from which only the relevant part for the filling process is included in figure\,\ref{fig:fill_circ}(a).

At $t = \SI{0.4}{h}$, the bypass valve is partially closed. This results in a decrease of the flow FIC550 and an increase in the bottom reboiler pressure PI530. After that, the compressor performance was raised and the flow FIC550 and PI530 further increase. At $t = \SI{2.1}{h}$, the bypass valve was fully closed leading to another flow reduction combined with an increase of $\mathrm{PI}530$ until it reaches a maximum of \SI{2.84}{bar} at $t = \SI{2.6}{h}$ and from here, staying constant. This is an indication for the first liquefaction of the GXe using the LXe at the top as cooling medium. A pressure difference of \SI{0.84}{bar} is required to initiate the liquefaction process given the constant pressure of PI510$= \SI{2}{bar}$ in the column.

At $t = \SI{3.05}{h}$, the LXe in the bottom vessel reaches the level meter $\mathrm{DP}540$. From here on, xenon was filled until $t = \SI{5.75}{h}$ with flows of up to \SI{15}{kg/h} (\SI{43}{slpm}). The total amount filled is $\mathrm{DP}540 = \SI{15.2}{mbar}$ corresponding to about \SI{33.8}{kg} of LXe.

After the filling, the compressor was stopped and the bottom reboiler was isolated from the column to monitor the containment/stability over the course of several hours before starting the circulation.

\subsection*{Internal liquid circulation}
The internal liquid circulation measurements including different flow values are presented in figure\,\ref{fig:fill_circ}(b): The top panel shows the column pressure PI510 (green), and the bottom reboiler pressure PI530 (red). The bottom panel contains the flow FIC550 (light blue) induced by the compressor.

The process was started at $t = \SI{0.35}{d}$ by partially opening the LXe bypass valve and starting the compressor. The pressure in the column was kept constant at $\mathrm{PI}510 = \SI{2.00(3)}{bar}$ with the help of the PID regulated heaters in the reboiler copper plate. Starting from $t = \SI{0.45}{d}$, a pressure increase to $\mathrm{PI}530 = \SI{3.01(5)}{bar}$ in the bottom reboiler vessel was required to fully liquefy the related xenon gas flow of \SI{36(1)}{kg/h} (\SI{103(3)}{slpm}) and to push the generated LXe from the reboiler bottom \SI{1.5}{m} upwards through the bypass valve back into the package tube's LXe feed port.

This flow configuration was kept for three days to monitor the stability of the RRS. Starting from $t = \SI{3.3}{d}$, the compressor performance was successively increased to reach larger flows. Between $\SI{3.6}{d} < t< \SI{4.3}{d}$, the RRS was performed stably at a flow of \SI{78(1)}{kg/h} (\SI{219(2)}{slpm}), slightly above the design target value. At $t = \SI{4.52}{d}$, a maximum operation flow of \SI{91(2)}{kg/h} (\SI{258(6)}{slpm}) at $\mathrm{PI}530 = \SI{3.25(1)}{bar}$ was achieved. This is about \num{1.3} times above the desired target flow and sufficient for an efficient radon removal in XENONnT. All relevant FIC550 flow values as well as reboiler bottom pressures PI530 are summarized in table \ref{tab:states2}.

\begin{table}[htbp]
\caption{Temperature and pressure values of the intermediate states during internal liquid circulation at a flow of \SI{91(2)}{kg/h} (\SI{258(6)}{slpm}). In order to describe the full cycle, some values with ($^{\ddag}$) are computed from measured values ($^{\dag}$), or are based on assumptions ($^{*}$).}
\centering
\begin{tabular*}{\columnwidth}{lccl}
\hline
 State & T [K] & P [bar] & Part\\ \hline
\textcircled{\raisebox{-0.3pt}{\scriptsize{1a}}} & 178.2$^{\ddag}$ & 2.03$^{\dag}$ & Reboiler top\\ 
\textcircled{\raisebox{-0.3pt}{\scriptsize{1b}}} & 181.3$^{\dag}$ & 2.03$^{\dag}$ & Top Condenser\\ 
\textcircled{\raisebox{-0.3pt}{\scriptsize{1c}}} & 268.8$^{\dag}$ & 2.03$^{\dag}$ & GXe/GXe HE to 4MP\\ 
\textcircled{\raisebox{-0.3pt}{\scriptsize{1d}}} & 285.2$^{\dag}$ & 1.58$^{\dag}$ & 4MP inlet\\ \hline
\textcircled{\raisebox{-0.3pt}{\scriptsize{2a}}} & 299.9$^{\dag}$ & 3.77$^{\dag}$ & 4MP outlet\\ 
\textcircled{\raisebox{-0.3pt}{\scriptsize{2b}}} & 218.1$^{\dag}$ & 3.77$^{*}$ & GXe/GXe HE to spiral\\ 
\textcircled{\raisebox{-0.3pt}{\scriptsize{2c}}} & 196.7$^{\dag}$ & 3.25$^{\dag}$ & Spiral\\ \hline
\textcircled{\raisebox{-0.3pt}{\scriptsize{3}}} & 188.5$^{\ddag}$ & 3.25$^{\dag}$ & Reboiler bottom\\ \hline
\textcircled{\raisebox{-0.3pt}{\scriptsize{4}}} & 178.2$^{\ddag}$ & 2.03$^{\dag}$ & Internal Bypass\\ \hline
\end{tabular*}
\label{tab:states}
\end{table}

In the following, the maximum flow measurement is used to describe the thermodynamic cycle. The temperature and pressure at each intermediate state of the Clausius-Rankine cycle are summarized in table \ref{tab:states}. All given pressure values are measured, except for state \textcircled{\raisebox{-0.3pt}{\scriptsize{2b}}}, where no pressure measurement is available. Therefore, this value is an assumption. Furthermore, some temperature values were derived from the measured pressure at a given state using the saturation properties. This is specifically the case for RTDs in states with LXe, where the limited contact between sensor and metal enclosure leads to temperature offsets towards warmer values.

\begin{figure*}[t]%
    \centering
    \subfloat[\centering $p$-$h$ diagram]{{	\includegraphics[width=0.47\textwidth]{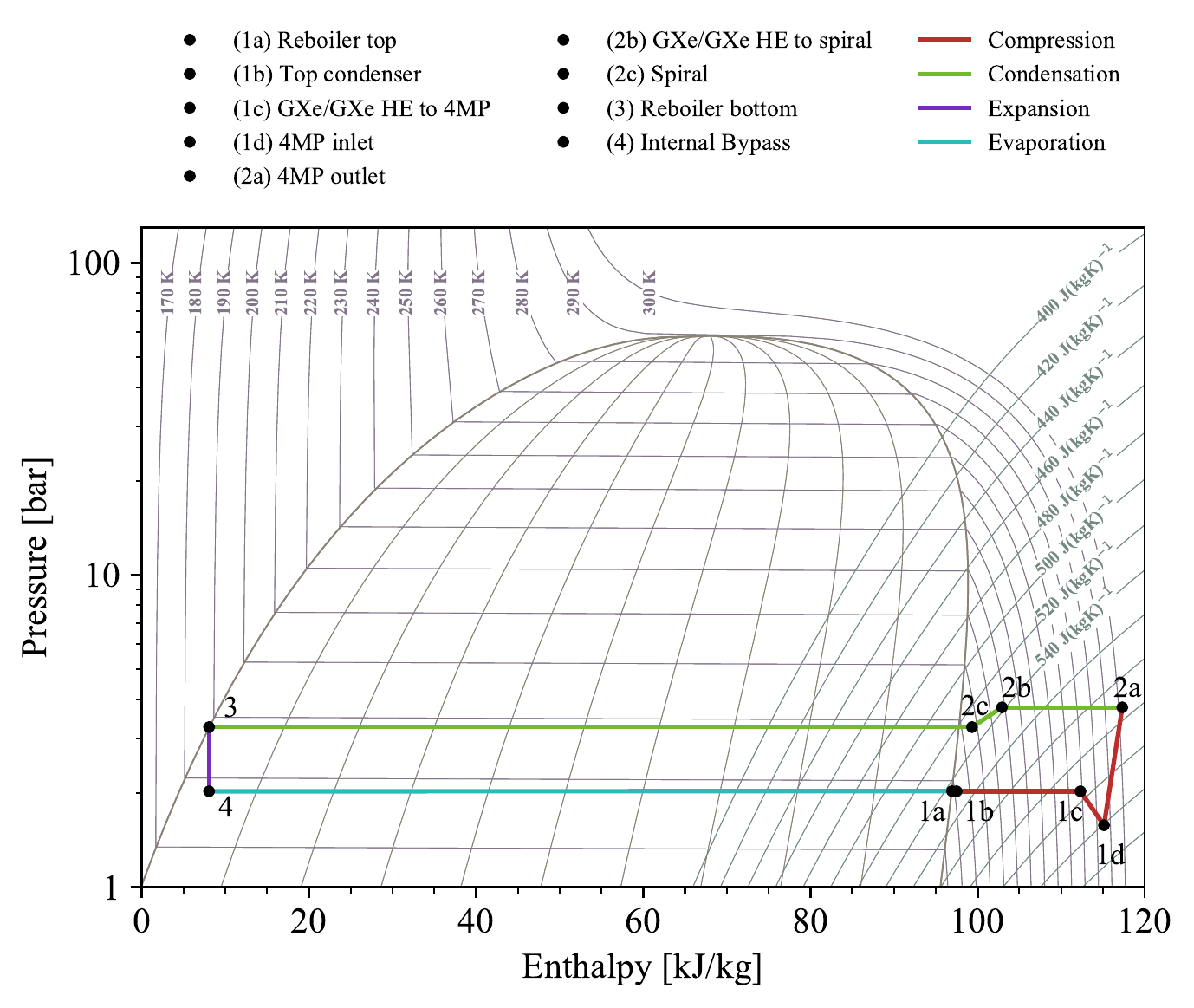} }}%
    \qquad
    \subfloat[\centering $T$-$s$ diagram]{{	\includegraphics[width=0.47\textwidth]{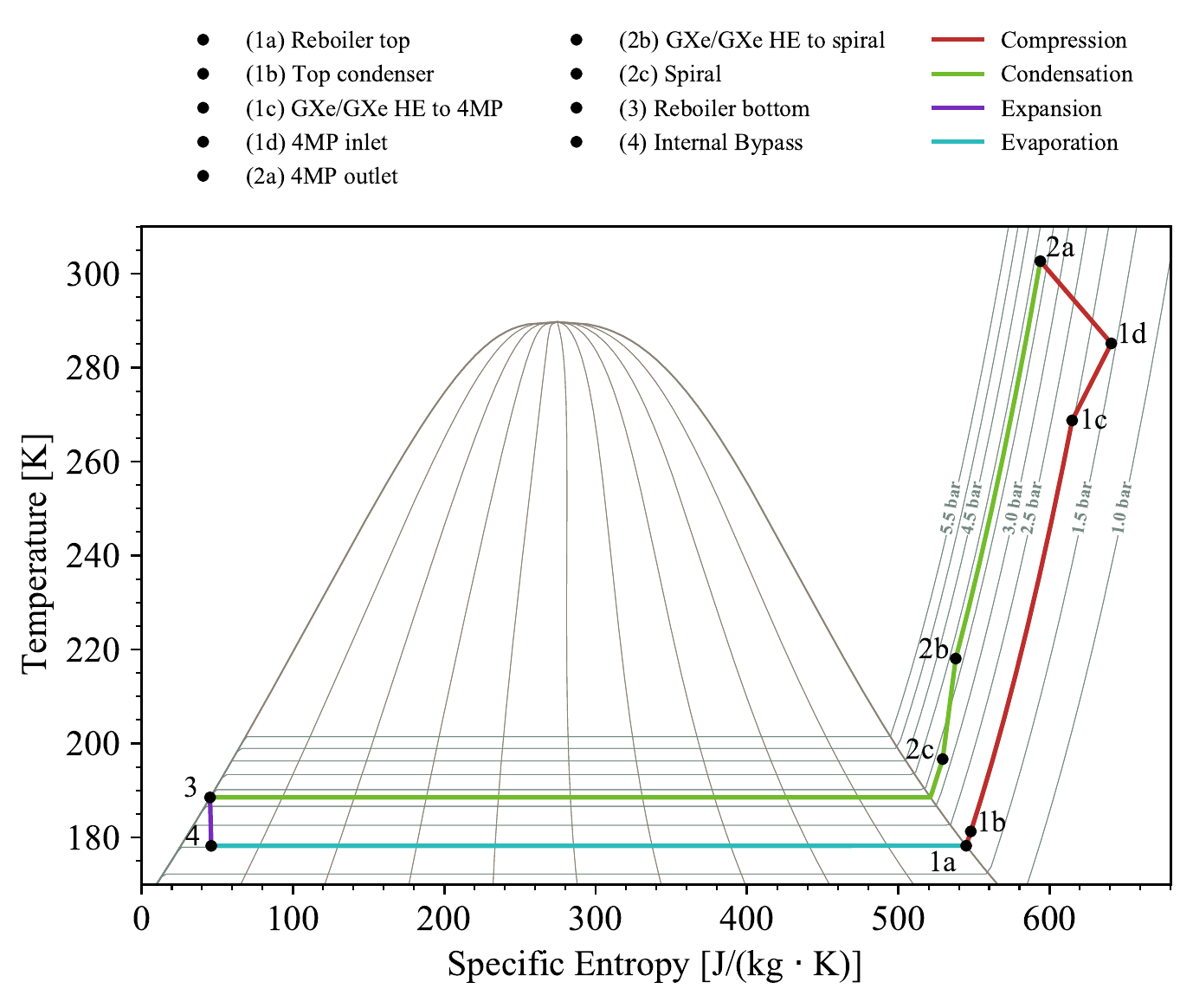}}}%
    \caption{The left-turning Clausius-Rankine cycle at a process flow of \SI{91.3}{kg/h} (\SI{258}{slpm}) with the four main states compression (red), condensation (green), expansion (purple), and evaporation (light blue) subdivided into intermediate states for a throttle efficiency $\epsilon_\mathrm{thr} = 1$. The $p$-$h$-diagram (a) indicates also lines of equal vapor quality ($f_\mathrm{vap}$), isoterms and isentropic lines. The $T$-$s$-diagram (b) indicates also lines of equal vapor quality ($f_\mathrm{vap}$), as well as isobaric lines. The different states were computed using the $CoolProp$ library \cite{cool}.}%
    \label{fig:ph_ts}%
\end{figure*}

Values for the enthalpy $h$ and specific entropy $s$ are computed from the temperature and pressure values at each state and are summarized in a $p$-$h$ diagram in figure\,\ref{fig:ph_ts}(a) and a $T$-$s$ diagram in figure\,\ref{fig:ph_ts}(b). In both diagrams, the compression is highlighted in red, the condensation in green, the expansion in purple, and the evaporation in light blue. Note that the expansion (\textcircled{\raisebox{-0.8pt}{3}}$\rightarrow$\textcircled{\raisebox{-0.8pt}{4}}) through the bypass valve (throttle) is an irreversible process. Therefore, the enthalpy stays constant as indicated by the vertical purple line in figure\,\ref{fig:ph_ts}(a), while the specific entropy increases as indicated by the purple line with a small inclination, not visible in figure\,\ref{fig:ph_ts}(b). This also shows that a fraction of the initially saturated LXe evaporates during the expansion. The distance from the saturated liquid curve to state \textcircled{\raisebox{-0.8pt}{4}} indicates the related vapor quality $f_\mathrm{vap}$ which indicates the gas fraction that is created. The remaining liquid fraction $f_\mathrm{liq}$ is then given by $(1-f_\mathrm{vap})$. In reality, the throttle efficiency $\epsilon_\mathrm{thr}$ is expected to be smaller than 1 depending on the size of heat losses across the throttle. This can lead to an increase in enthalpy between the states \textcircled{\raisebox{-0.8pt}{3}} and \textcircled{\raisebox{-0.8pt}{4}}. 

\begin{table*}[t]
\setlength{\tabcolsep}{2.3pt}
\centering
\caption{Parameters during internal liquid circulation. Measured variables are indicated with ($^{\dag}$), while computed values are indicated by ($^{\ddag}$). The value of $f_\mathrm{liq}$ is calculated for a perfect throttle $\epsilon_\mathrm{thr}=1$ (as shown in figure \ref{fig:ph_ts}a), while $\epsilon_\mathrm{thr}$ is calculated using equation (\ref{eq:throttle}).}
\begin{tabular*}{0.72\linewidth}{cccccc}
\hline
 ${F^{\dag}}$ [kg/h] ([slpm]) &PI530$^{\dag}$ [bar] & $Q_\mathrm{B'}^{\ddag}$ [W] & $Q_\mathrm{H}^{\dag}$ [W] & $f_\mathrm{liq}^{\ddag}$ & $\epsilon_\mathrm{thr}^{\ddag}$ [\SI{90}{\percent} C.L.]\\ \hline
    0      & $2.87  \pm 0.01$  & 0 & $1129 \pm \, 6$ & - & -\\
 $33.6 \pm 1.4$  \, ($95  \pm 4$)     & $3.05  \pm 0.01$  &  \, $841  \pm 36$& $1086 \pm 10$& $ 0.9658 \pm 0.0005$ & $> 0.94$\\
 $49.2 \pm 1.8$ ($139 \pm 5$)   & $3.08  \pm 0.01$    & $1230 \pm 44$   & $1066 \pm 12$ & $0.9649 \pm 0.0005$ &  $> 0.94$\\
 $55.6  \pm 1.8$ ($157 \pm 5$)  & $3.10  \pm 0.01$    & $1389  \pm 45$  & $1067 \pm 13$ & $0.9644 \pm 0.0005$ & $> 0.95$\\
 $63.0  \pm 1.8$ ($178 \pm 5$)   & $3.12  \pm 0.01$    & $1574  \pm 45$  & $1060 \pm 13$ & $0.9638   \pm 0.0005$ & $> 0.95$\\
 $68.7 \pm 1.8$ ($194 \pm 5$)   & $3.14  \pm 0.01$    &  $1714  \pm 44$ & $1060 \pm \, 8$ & $0.9632 \pm 0.0005$ & $> 0.96$\\
 $77.5 \pm 1.8$ ($219 \pm 5$)   & $3.20  \pm 0.01$    & $1933  \pm 44$  & $1048 \pm 11$ & $0.9616 \pm 0.0005$ & $> 0.96$\\
 $86.7 \pm 1.8$ ($245 \pm 5$)    & $3.22  \pm 0.01$   &  $2162  \pm 43$ & $1032 \pm 12$ & $0.9611 \pm 0.0005$ & $> 0.96$\\
 $91.3 \pm 2.1$ ($258 \pm 6$)   & $3.25  \pm 0.01$    &  $2274  \pm 52$  & \, $991 \pm 17$ & $0.9602 \pm 0.0005$ & $> 0.94$\\\hline
\end{tabular*}
\label{tab:states2}
\end{table*}

The throttle efficiency is calculated in the following for the different flow steps presented in table \ref{tab:states2} to further evaluate the RRS performance. For small heat losses across the bypass valve, the efficiency $\epsilon_\mathrm{thr}$ is expected to be close to \num{1}. The assumption is made that the throttle is made of the LXe bypass valve including the stainless steel tubes from the reboiler bottom vessel up to the LXe feed port. Furthermore, the external heat input is expected to be constant during all measurements presented and can therefore be neglected for the throttle efficiency calculations. The top condenser's cooling power $Q_\mathrm{L}$, and thus, the reflux liquid flow $L$, was kept constant throughout all measurements. As described in section \ref{sec:radon_column}, the additional electrical heating power $Q_\mathrm{H}$ is applied to the reboiler copper plate to balance the LXe flow $L$ coming from the top condenser. The presented values $Q_\mathrm{H}$ were measured once both LXe levels inside the bottom reboiler part as well as inside the top part were constant after a flow change.

The LXe flow $ L^\prime$ entering the top reboiler vessel is then given by
\begin{equation}
    L^\prime = L + qF,
\end{equation}
with $q = \epsilon_\mathrm{thr} \cdot f_\mathrm{liq}$. Due to the mass balance in this location, it follows
\begin{equation}
\label{eq:balance}
    L^\prime =  V^\prime,
\end{equation}
with $V^\prime$ being the evaporated GXe leaving the top reboiler.
The related heat transfer can be further split into different contributions:
\begin{align}
    Q_{L^\prime} &= Q_{L} + Q_{qF} = Q_{B^\prime} + Q_{H} + Q_{S}.
\end{align}

Here, $Q_{qF}$ is the stored energy from the flow fraction $qF$, $Q_{B^\prime}$ is the heat input transferred through the reboiler copper plate from the bottom to the top part, and $Q_{S}$ is the heat input from the spiral immersed in LXe of the top reboiler part. The different $Q$ terms are given by
\begin{align}
    Q_{L}  &=  \left.Q_\mathrm{H}\right|_{\SI{0}{kg/h}} = const., \\
    Q_{qF} &= \epsilon_\mathrm{thr} \cdot f_\mathrm{liq} \cdot F \cdot \left.\Delta H_\mathrm{vap}^\mathrm{Xe}\right|_{\mathrm{PI510}}, \\
    Q_{B^\prime} &= F \cdot \left.\Delta H_\mathrm{vap}^\mathrm{Xe}\right|_{\mathrm{PI530}}, \\
    \label{eq:q_spiral}
    Q_{S}  &=  F \cdot c_\mathrm{p} \cdot \Delta T_\mathrm{S},
\end{align}
Here, $\left.Q_\mathrm{H}\right|_{\SI{0}{kg/h}}$ is the applied heating power at zero flow $F$, $\left.\Delta H_\mathrm{vap}^\mathrm{Xe}\right|_{\mathrm{PI510}}$ is the enthalpy of evaporation for the given column pressure $\mathrm{PI510}$. The pressure in the column was kept constant throughout all measurements at a value of $\mathrm{PI510} = \SI{2.00(1)}{bar}$. $\left.\Delta H_\mathrm{vap}^\mathrm{Xe}\right|_{\mathrm{PI530}}$ is the enthalpy of evaporation for the given pressure $\mathrm{PI530}$ in the bottom reboiler part. Its value depends on the flow and the values are summarized in table \ref{tab:states2}.The specific heat capacity of the GXe inside the spiral is given by $c_\mathrm{p}$, while the temperature difference across the spiral is $\Delta T_\mathrm{S}$. The applied heating power $Q_{H}$ is used to achieve mass balance as described above and is measured for each given flow. Using equations (\ref{eq:balance}) to (\ref{eq:q_spiral}) and solving for the throttle efficiency $\epsilon_\mathrm{thr}$ leads to
\begin{equation}
\label{eq:throttle}
     \epsilon_\mathrm{thr} =  \frac{F \cdot \left( \left.\Delta H_\mathrm{vap}^\mathrm{Xe}\right|_{\mathrm{PI530}} + c_\mathrm{p} \cdot \Delta T_\mathrm{S}\right) + \Delta Q_{H}}{ f_\mathrm{liq} \cdot F \cdot \left.\Delta H_\mathrm{vap}^\mathrm{Xe}\right|_{\mathrm{PI510}}},
\end{equation}
with $\Delta Q_{H} =  Q_{H} - \left.Q_\mathrm{H}\right|_{\SI{0}{slpm}}$.

The temperature difference $\Delta T_\mathrm{S}$ cannot be accurately measured as the connection between the temperature sensors before and after the spiral to the pipe introduce unknown systematics. This also does not allow to determine the temperature at which the GXe enters the bottom reboiler. However, following equation (\ref{eq:throttle}), the conservative assumption can be made that the spiral has no effect ($c_\mathrm{p} \cdot \Delta T_\mathrm{S} = 0$) and that the GXe enters the reboiler at saturation temperature. This yields the lowest value for the efficiency and a lower limit can be derived. Taking the uncertainties for the measured flow, pressure and heating powers into account, a throttle efficiency of $\epsilon_\mathrm{thr} > 0.94 \left( \SI{90}{\percent}\, \mathrm{C. L.}\right)$ is calculated for the maximum flow of $F = \SI{91.3(21)}{kg/h} (\SI{258(6)}{slpm})$. The efficiency limits for the remaining flows are summarized in table \ref{tab:states2} and all yield similar results close to one as expected for small heat losses across the bypass valve and the connecting tubes.

The reboiler performance itself is investigated for the maximum flow. The flow was fully liquefied exiting from the bottom reboiler vessel. This corresponds to a heat transfer from the bottom to the top reboiler vessel through the copper plate of
\begin{equation}\label{QB}
     \left.Q_{B^\prime}\right|_{\SI{91}{kg/h}}= (2274\pm 52)\,\mathrm{W},
\end{equation}
with $\left.\Delta H_\mathrm{vap}^\mathrm{Xe}\right|_{\mathrm{3.25\,bar}}=89.70$\,kJ/kg \cite{NST} being the enthalpy of evaporation at the related pressure in the reboiler bottom part. Note that also here the conservative assumption was made that saturated GXe enters the reboiler for liquefaction after leaving the spiral. The heating power at the given flow was measured to be
\begin{equation}
    \label{QR}
     \left.Q_\mathrm{H}\right|_{\SI{91}{kg/h}}= \SI{991(17)}{W}.
\end{equation}
This leads to a total heat transfer of 
\begin{align}\label{Qtot}
    \left.Q_\mathrm{tot}\right|_{\SI{91}{kg/h}}=Q_{B^\prime}+ Q_\mathrm{H} = \SI{3265(69)}{W}.
\end{align}

From the heat transfer model for the reboiler based on conservative assumptions in Ref. \cite{k}, a maximum total heat transfer of \SI{2627}{W} is expected for the given pressure of PI$530 = \SI{3.25}{bar}$. The measured value is about \num{1.25} times larger, indicating that the liquefaction process is more efficient than expected. 

Finally, the coefficients of performance $COP_\mathrm{R}$ for the refrigerator part and  $COP_\mathrm{HP}$ for the heat pump part \cite{COP1} can be derived from the p-h-diagram. A range for $0.94 < \epsilon_\mathrm{thr} \leq 1$ is computed for the maximum flow:
\begin{align}
COP_\mathrm{R} &= \frac{h_\mathrm{1a} - h_4}{h_\mathrm{2a} - h_\mathrm{1a}} = 4.10-4.36, \\
COP_\mathrm{HP} & = \frac{|h_3 - h_\mathrm{2a}|}{h_\mathrm{2a} - h_\mathrm{1a}} = 5.36, 
\end{align}
with $h_\mathrm{i}$ being the enthalpy at each state. Traditional heat pumps feature COP values of 2-5 \cite{COP}, indicating that the RRS performance is on the upper end, and thus, highly efficient.
\section{Conclusion}
\label{sec:conclusion}
A radon removal system based on cryogenic distillation was built to reduce the radon background in XENONnT. The derived radon removal strategy based on the radon source locations shows that a reduction by a factor two for type 1 radon sources within the detector volume can be achieved for a process flow of \SI{62}{kg/h} (\SI{175}{slpm}) given the \SI{8.4}{tonne} xenon inventory. An additional factor two is expected when type 1b radon sources from the detector's GXe volume are extracted and converted into type 2 sources before they can enter the detector's LXe. 

The distillation process was designed using the \linebreak McCabe-Thiele method, where a modification was \linebreak needed: Instead of a physical xenon extraction flow at the bottom that removes radon from the system, the radon decay itself was added as an equivalent extraction flow. This makes it possible to describe the distillation process, while conserving the required particle mass balance inside the column. It was shown, that the energy efficient design based on a Clausius-Rankine cycle with phase-changing cooling medium including a custom-made heat pump drastically reduces the externally applied cooling power.

A first liquefaction of radon-depleted xenon in the bottom vessel of the reboiler heat exchanger using the liquid xenon in the top vessel was achieved. A heat transfer of \SI{3265(69)}{W} was demonstrated, a factor 1.25 larger than modeled in \cite{k}. The system was operated in an internal liquid circulation cycle using a LXe bypass valve. A stable flow of $(91\pm2)$\,kg/h (($258\pm 6$)\,slpm) was achieved, a factor 1.3 larger than required by design. In this test, the efficiency of our LXe bypass throttle valve was evaluated to be $\epsilon_\mathrm{thr} > 0.94  \left( \SI{90}{\percent}\, \mathrm{C. L.}\right)$.

The achieved process flow of the LXe extraction mode in combination with the GXe extraction mode leads to an expected total radon reduction by a factor of 4.7. With an initial activity concentration of $4.25\,\upmu$Bq/kg in XENONnT, a final \isotope[222]{Rn} activity concentration of $<1\,\upmu$Bq/kg is expected.

\begin{acknowledgements}
We thank the XENON Collaboration and LNGS staff, especially D. Tatananni and R. Corrieri, for supporting the installation of the system at LNGS.
We thank the group of M. Lindner at MPIK Heidelberg for determining the radon emanation of all components of the removal system. We acknowledge the work of the Precision Mechanical and the Electrical Workshop of the Institute for Nuclear Physics at M\"unster University.
The development of the system at Muenster University has been supported by the German Ministery for Education and Research BMBF (05A17PM2, 05A20PM1).
\end{acknowledgements}

\bibliographystyle{aip2}

\end{document}